\documentclass[aps,prx,twocolumn,superscriptaddress,floatfix,noeprint]{revtex4-2}
\usepackage[utf8]{inputenc}
\usepackage[T1]{fontenc}
\usepackage{amssymb}
\usepackage{graphics}
\usepackage{float}
\graphicspath{{figures/}}
\usepackage{mathtools}
\allowdisplaybreaks
\usepackage{braket}
\usepackage{bm}
\usepackage{enumitem}
\usepackage[section]{placeins}
\usepackage{color}
\usepackage{siunitx}
\usepackage{chngcntr}
\usepackage{microtype}

\usepackage[colorlinks=true,allcolors=blue]{hyperref}

\DeclarePairedDelimiter\abs{\lvert}{\rvert}
\DeclarePairedDelimiter\mean{\langle}{\rangle}
\DeclarePairedDelimiter\norm{\lVert}{\rVert}
\DeclareMathOperator{\tr}{tr}
\DeclareMathOperator{\var}{var}
\newcommand*{\up}{\uparrow}
\newcommand*{\down}{\downarrow}
\newcommand*{\diss}{\mathcal{D}}
\newcommand*{\EPR}{\Delta_\mathrm{EPR}}
\newcommand*{\s}[2]{\sigma^{#1}_{#2}}
\newcommand*\isotope[2]{\textsuperscript{#2}#1}

\usepackage[normalem]{ulem}

\definecolor{gray}{rgb}{.6,.6,.6}
\definecolor{purple}{rgb}{.6,.1,.6}

\begin{document}

%\preprint{}

\title{Entangling nuclear spins in distant quantum dots via an electron bus}


\author{Miguel Bello}
\email{miguel.bello@mpq.mpg.de}
\affiliation{Instituto de Ciencia de Materiales de Madrid (ICMM-CSIC), ES-28049 Madrid, Spain}
\affiliation{Max Planck Institute of Quantum Optics, DE-85748 Garching, Germany}

\author{Mónica Benito}
\affiliation{Department of Physics, University of Basel, CH-4056 Basel, Switzerland}

\author{Martin J. A. Schuetz}
\thanks{This work was done prior to joining AWS}
\affiliation{Amazon Quantum Solutions Lab, Seattle, Washington 98170, USA}
\affiliation{AWS Center for Quantum Computing, Pasadena, California 91125, USA}

\author{Gloria Platero}
\affiliation{Instituto de Ciencia de Materiales de Madrid (ICMM-CSIC), ES-28049 Madrid, Spain}

\author{Géza Giedke}
\affiliation{Donostia International Physics Center (DIPC), ES-20018 Donostia-San Sebastián, Spain}
\affiliation{IKERBASQUE,  Basque  Foundation  for  Science,  ES-48013,  Bilbao,  Spain}

%\thanks{}
%\affiliation{}


\date{\today}

\begin{abstract}
  We propose a protocol for the deterministic generation of entanglement
  between two ensembles of nuclear spins surrounding two distant quantum dots. The 
  protocol relies on the injection of electrons with definite polarization 
  in each quantum dot and the coherent transfer of electrons from one 
  quantum dot to the other. 
  Computing the exact dynamics for small systems, and using an effective
  master equation and approximate non-linear equations of motion for larger
  systems, we are able to confirm that our protocol indeed produces entanglement
  for both homogeneous and inhomogeneous systems. Last, we analyze the 
  feasibility of our protocol in several current experimental platforms.
  %
  % Using an effective master equation, and 
  % approximate non-linear equations of motion for certain spin correlations, 
  % we are able to asses the amount of entanglement generated in homogeneous 
  % as well as inhomogeneous systems, where each individual nuclei couples 
  % differently to the electrons. Last, we analyze the feasibility of our 
  % protocol in several current experimental platforms. Our findings indicate
  % that the proposed protocol allows the deterministic generation of 
  % entanglement even in the case of moderate inhomogeneity.
\end{abstract}

%\keywords{}

\maketitle

\section{Introduction \label{sec:intro}}
Electron spin coherence in quantum dots is affected by several phenomena, including hyperfine interaction with surrounding nuclei \cite{schliemann2003,inarrea2007,chekhovich2013}. Scientists have developed various techniques (spin echo, nuclear polarization or narrowing, etc.) to reduce the effects of hyperfine interaction \cite{bluhm2011,MM+Marcus16}, or to get rid of the nuclear spins altogether \cite{schreiber2014}. Going the opposite way, different works have explored the idea of harnessing the long coherence times (recently demonstrated for QDs experimentally \cite{Waeber2016,Gangloff2019}), and slow and rich dynamics of the nuclear spin ensembles \cite{SiLo07,DVK+Nazarov09,RuLe10,KMA+I10,rudner2011,Kessler2012a,RuLe13,LL+Platero13,KA+Tarucha15,Shumilin21}, which make them good candidates for long-time quantum information processing and storage platforms (demonstrated both theoretically \cite{dobrovitski2006,witzel2007,KST+Fl09,Gangloff2019,denning2019} and experimentally \cite{taylor2003,hensen2020}). %In this work we also view the nuclear spins as a resource rather than an obstacle, and propose a protocol to produce entanglement between two nuclear ensembles deterministically.

In this work we follow this research line, and aim to design a protocol that produces entanglement between two distant nuclear ensembles deterministically. This has been pursued in Refs.~\cite{schuetz2013,schuetz2014} for the two nuclear ensembles surrounding a double quantum dot. For an analogous operation at longer distances, as explored in Refs.~\cite{benito2016,yang2016} for entanglement generation between distant electrons, one may use a coherent electron bus, such as a longer array of quantum dots \cite{baart2016,ban2019,mills2019,yoneda2020}, a chiral quantum-Hall-effect channel \cite{benito2016}, or surface acoustic waves \cite{bauerle2018,takada2019,JM+Meunier20}. The idea behind most of these proposals is that the moving qubits (a current of electrons) act as a joint bath, whose overall effect is to induce the decay of the system towards a target steady state. By engineering the bath we can modify the properties of this steady state, in particular, we can ensure that the two parties (the two electrons or two ensembles of nuclear spins) are entangled. 

Our work comprises the first proposal for entanglement generation between \emph{distant} ensembles of nuclear spins connected via an electron bus. Two different regimes are of potential interest to apply our scheme: On the one hand, the large nuclear ensembles that surround QDs fabricated with GaAs, InGaAs, and \isotope{Si}{29}-rich Si, with a number of spins ranging from 100 to $10^6$ \cite{jackson2020}; on the other, nuclear ensembles with just a few nuclei ($<10$)  present, for example, in purified Si \cite{hensen2020,asaad2020}. We also discuss how to detect the entanglement experimentally using different entanglement witnesses suitable for each case.

% One important difference as compared with analogous proposals for quantum optical systems \cite{muschik2011}, is that solid state devices are in general more noisy and inhomogeneous than their quantum optical counterparts. This makes it challenging to apply ideas from the realm of quantum optics to solid-state systems. Yet, as we will show, our protocol is able to produce entanglement in both homogeneous and inhomogeneous settings, and since it is a dissipative protocol, it is robust against noise and imperfections in its execution. 
% Recent experimental developments \cite{?} suggest that our proposal could 
% be implemented in the near term.

% \Comment{GG: actually, I don't think there's a
%   collective enhancement: the coupling is stronger, if
%   the QD is smaller: if the QD was as small as one nucleus, we'd have
%   $g=A_H$; for $N$ homogeneously coupled nuclei: $g=A_H/\sqrt{N}$;
%   there \emph{is} an enhancement if you keep the QD fixed and increase
%   the number of coupled nuclei} 

% \Comment{MB: One could argue  that there is an enhancement in the 
% entanglement produced as we increase the number of nuclei in each ensemble}

\section{Design of the protocol \label{sec:protocol}}

\subsection{System}

\begin{figure}
  \includegraphics{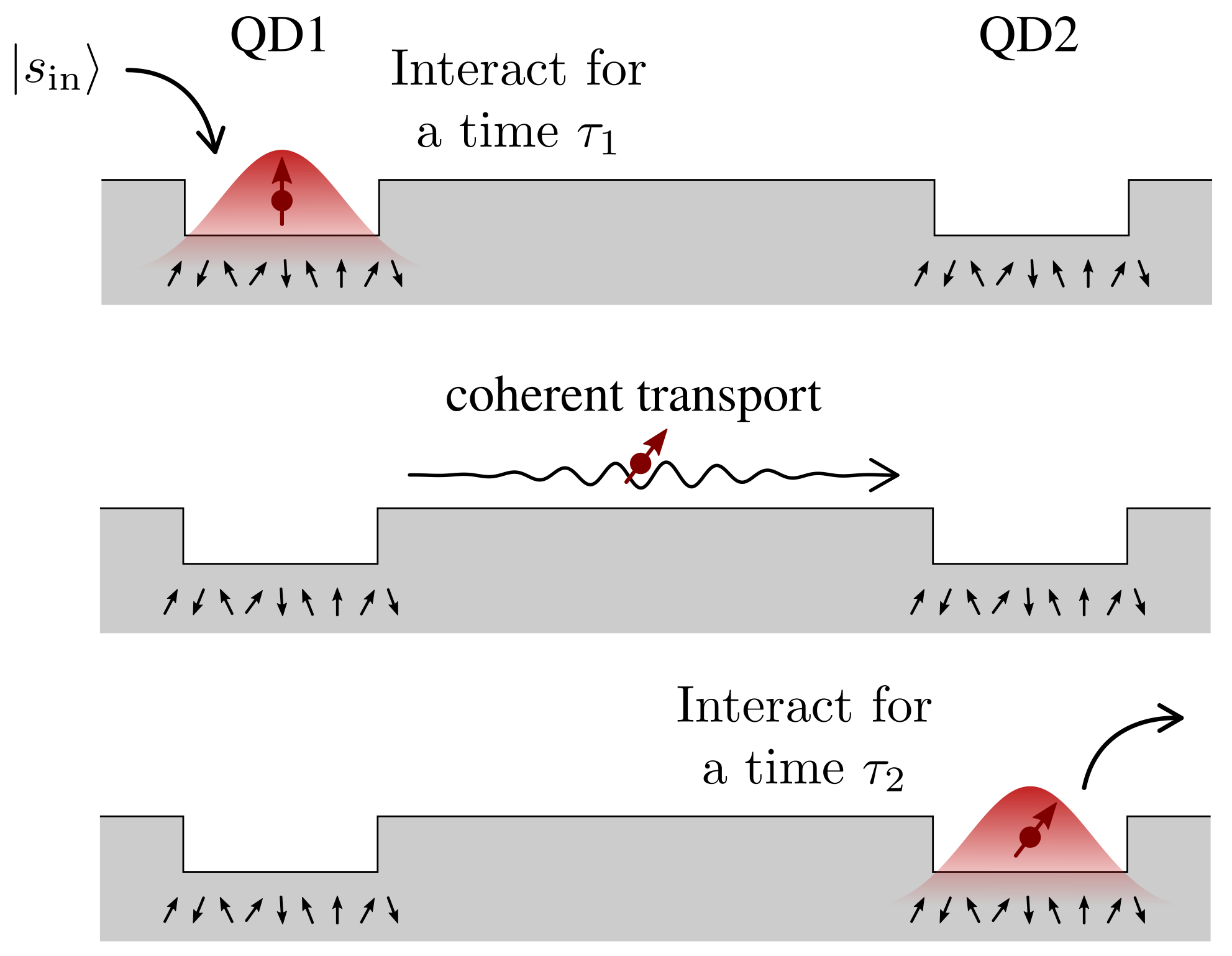}
  \caption{Schematics of the system under consideration and the manipulations performed in each step of the protocol. \label{fig:schematics}}
\end{figure}

Our protocol is designed for a device such as the one shown schematically in Fig.~\ref{fig:schematics}: Two QDs, henceforth labelled as QD1 and QD2, are connected to electron reservoirs so that electrons with definite polarization can be injected into each QD, and they can be released from either QD at will. Furthermore, the two QDs are connected with each other via a coherent electron bus, which can transport electrons from QD1 to QD2.

If QD$j$ is occupied by an electron, the nuclei in a small volume around it 
will interact with it via the hyperfine interaction. We consider the Fermi 
contact interaction between an s-type conduction band electron and the 
nuclear spins, which is given by the Hamiltonian~\cite{schliemann2003}
\begin{equation}
  \begin{split}
    H_j & = g_j \bm{A}_j \cdot \bm{S}_j\\ & = 
    g_j A^z_j S^z_j + \frac{g_j}{2}\left(A^+_j S^-_j + A^-_j S^+_j\right)
    \equiv H^z_j + H^\perp_j\,.
  \end{split}
\end{equation} 
Here, $\bm{S}_j$ denotes the electronic spin, while 
$\bm{A}_j = \sum_i g_{ij} \bm{I}_{ij}$ is a collective nuclear spin 
operator; $\bm{I}_{ij}$ is the spin of the $i$th nucleus in QD$j$. The 
ladder operators are defined as usual $S^\pm_j=S^x_j \pm iS^y_j$ 
(analogously for $A^\pm_j$). 
We consider spin-$I$ nuclei, and a total of $N_j$ nuclei in each ensemble. The 
dimensionless coupling constants $g_{ij}$ may differ depending on the 
electronic wavefunction at the positions of the nuclei and the nuclear species, 
they are normalized such that $\sum_i g_{ij}^2 = 1/(2I)$ \footnote{Usually, the contact hyperfine interaction in quantum dots is written as $H=A_H\sum_i \alpha_i \bm{I}_i \cdot \bm{S}$, where $\alpha_i$ gives the relative coupling to the $i$th nucleus, $\sum_i \alpha_i = 1$. Instead, we use an equivalent parametrization of the hyperfine interaction in terms of collective spin operators $A^\pm_j$, which satisfy approximately bosonic commutation relations $\mean{[A^-_j,A^+_j]}\approx (-1)^j$ for highly-polarized nuclear states. This simplifies the subsequent analysis of the protocol based on the Holstein-Primakoff approximation}. The constants 
$g_j=A_H/\sum_i g_{ij}$ set the overall strength of the hyperfine 
interaction in each QD; $A_H$ is a constant that depends on the material 
and the density (concentration) of nuclei around the QD 
\cite{schliemann2003,philippopoulos2020}. 

Within this model there are implicit several approximations: We assume that the electron injection/retrieval process is fast on the scale of the hyperfine interaction, so that $H_j$ is switched on/off abruptly whenever an electron is added to/removed from QD$j$, and that it is spin-independent \cite{takada2019}. We also neglect other effects, such as non-contact terms of the hyperfine interaction and nuclear dipole-dipole interactions, which usually are much weaker compared with the contact interaction \cite{coish2009}. We also neglect spin-orbit coupling, which is weak in GaAs, and all electron-spin decoherence processes. These simplifications are discussed and justified in Section~\ref{sec:experiment}.

\subsection{Protocol}

The protocol 
% is based on the analogy between the collective nuclear spin-excitation the QD electrons couple to and a bosonic mode, and related ideas in quantum optics \cite{muschik2011,Krauter10}. It 
consists of a sequence of interactions that generate a dissipative dynamics whose principal (Lindbladian) terms stabilize a spin-squeezed state of the two ensembles \cite{muschik2011,Krauter10}. The interactions are designed such that the main unwanted dephasing terms cancel out.
% In the following we introduce the electronic transport protocol that induces
% correlations between the two distant nuclear spin ensembles, and the 
% experimental measurements whereby entanglement between the two nuclear
% ensembles can be detected. The protocol is based on related ideas in 
% quantum optics \cite{muschik2011,Krauter10} and the analogy between the 
% collective nuclear spin-excitation the quantum dot electron couples to and 
% a bosonic mode. Our protocol consists of a sequence of interactions that 
% generate a dissipative dynamics whose principal (Lindbladian) terms 
% stabilize a spin-squeezed state of the two ensembles. The interactions are designed such that 
% the main unwanted dephasing terms cancel out. 
Specifically, the protocol consists in the repetition of a cycle comprised 
of $2+2k$ steps ($k\geq 1$). At the beginning of each step, a new electron with definite
polarization is injected into QD1. This electron interacts with the nuclei 
in QD1 for a certain period of time $\tau_1$ and then it is transported to 
QD2, where it interacts with the nuclei there for a period $\tau_2$, see 
Fig.~\ref{fig:schematics}. At the end of each step the electron in QD2 is
released from the device, leaving it ready to start a new step. The dwell times 
$\tau_1$ and $\tau_2$, and the polarization of the injected electrons at 
each step are required to fulfill a certain pattern: in the first step (injected spin ``up'') $g_1\tau_1=\mu$ and $g_2\tau_2=\nu$, with $\tau_1>\tau_2$ ($\abs{\mu}>\abs{\nu}$), while in the second step (injected spin ``down'') $g_1\tau_1=\nu$ and $g_2\tau_2=\mu$. In the remaining $2k$ steps each nuclear ensemble evolves independently under the influence of $k$ consecutive electrons, all with the same polarization (``down'' for QD1 and ``up'' for QD2), that spend each a short time $\tau_j\propto k^{-1}$ trapped in QD$j$. All this information is summarized in Table~\ref{tab:protocol}.

% The rationale behind it is as follows: For short dwell times, each 
% electron acts like a Markovian environment for the nuclear ensembles. If 
% the electrons are transported coherently between QD1 and QD2 in the first 
% two steps of each cycle, there is an interference between the 
% ``flipped in QD1'' and ``flipped in QD2'' paths. Tracing out the initially 
% $z$-polarized electrons produces decoherence given by $A^\pm_j$ and 
% $A^z_j$ terms; since the unwanted $z$-terms act like an effective magnetic 
% field we can cancel them with additional local terms. These are produced
% in the remaining steps of each cycle, where each nuclear ensemble evolves 
% independently under the influence of $k$ consecutive electrons, all with 
% the same polarization, that each spend a short time $\tau_j\propto k^{-1}$ trapped in QD$j$.

The rationale behind this particular choice of the dwell times and injected electron states is the following: For short dwell times, we can neglect higher-order terms in the hyperfine interaction and each electron flips at most once when interacting with the nuclear ensembles. In the first two steps of the protocol, each electron is transported coherently between QD1 and QD2, such that interference between the ``flipped in QD1'' and ``flipped in QD2'' paths occurs. This uncertainty regarding where the electrons have flipped will be mapped onto an entangled nuclear state after tracing out (discarding) the electrons. Simultaneously, these two steps produce unwanted dynamics due to the $zz$ hyperfine terms. terms. On the nuclei, these latter terms act like an (in general inhomogeneous) effective magnetic field (the Knight field) that can be cancelled with additional local terms, which are produced in the remaining steps of each cycle. 

\begin{table}
  \caption{Electronic polarizations and dwell times for the steps that make 
  up a single cycle of the protocol. The third and fourth rows are steps 
  repeated a total of $k$ times each.  \label{tab:protocol}}
  \begin{ruledtabular}
  \begin{tabular}{cccc}
    step(s) & $\ket{s_\mathrm{in}}$ & $g_1 \tau_1$ & $g_2 \tau_2$ \\[0.2em] \hline
    $1$ & $\ket{\up}$ & $\mu$ & $\nu$ \rule{0pt}{1em}\\
    $2$ & $\ket{\down}$ & $\nu$ & $\mu$ \\
    $3, \dots, 2 + k$ & $\ket{\down}$ & $(\mu -\nu)/k$ & 0 \\
    $3 + k, \dots, 2 + 2k$ & $\ket{\up}$ & 0 & $(\mu - \nu)/k$ \\
  \end{tabular}
  \end{ruledtabular}
\end{table}

This specific pattern for the dwell times may be hard to fulfil in actual experiments if the coupling constants $g_j$ are not well known. However, the protocol also works if the pattern is satisfied only approximately. We are also assuming a perfectly coherent electron transfer from QD1 to QD2, but this again is not strictly necessary. For the sake of clarity here we will focus on the ``ideal'' implementation of the protocol, as given in Table~\ref{tab:protocol}, while the results including imperfect dwell times and electron transfer are shown in the Supplementary Material. 

The induced nuclear dynamics is Markovian by construction, 
since discarding the electron after each step results in a ``memoryless'' environment for the nuclei. According to the process described, the nuclei's reduced density matrix at the end of the $n$th step $\rho_n$, only depends on the nuclei's reduced density matrix at the end of the previous step,
\begin{equation}
  \rho_n = \tr_\mathrm{el} \left( e^{\tau_2 \mathcal{L}_2}
  \left[ e^{\tau_1 \mathcal{L}_1} [\chi_{n-1}] \right]
  \right). \label{eq:exactevol}
\end{equation}
In this equation $\chi_{n-1} = \rho_{n-1} \otimes \ket{s_\mathrm{in}}\bra{s_\mathrm{in}}$ is 
the density matrix of the entire system at the beginning of the step, and $\ket{s_\mathrm{in}}$ is the state of the injected 
electron. The Liouvillians that appear in it correspond to the coherent evolution given by 
the hyperfine interaction $\mathcal{L}_j \left[\chi\right] = 
-i \left[ H_j, \chi \right]$. If the dwell times are small, 
$\tau_j < \norm{H_j}^{-1}$ ($\tau_j< 2 I^{-1}A_H^{-1}$), we can approximate the 
dynamics by truncating the Taylor expansion of the exponentials up to 
second order. Then, the change in the density matrix after a single cycle 
$\Delta\rho$ is given by (see Supp.\ Material)
\begin{multline}
  \Delta\rho \simeq \frac{(\mu-\nu)^2}{4k}
  \left\{\mathcal{D}(A^-_1) + \mathcal{D}(A^+_2)\right\} [\rho] 
  \\
  + \frac{1}{4}\left\{\mathcal{D}(\mu A^+_1 + \nu A^+_2)
  + \mathcal{D}(\nu A^-_1 + \mu A^-_2)\right\} [\rho] \,,
  \label{eq:mastereq}
\end{multline}
where $\mathcal{D}(X)[\rho]\equiv X\rho X^\dagger - \{X^\dagger X,\rho\}/2$.

\subsection{Entanglement criteria}

To detect entanglement between the two nuclear ensembles in an experiment one can employ a spin-squeezing criterion \cite{muschik2011,schuetz2013} based on the following principle \cite{raymer2003}: Given two observables $X_j$ and $Y_j$ of each ensemble $j=1,2$ satisfying the commutation relation 
$[X_i,Y_j]=\delta_{ij}Z_j$, and linear combinations 
$u=\alpha X_1 + \beta X_2$ and $v=\alpha Y_1 - \beta Y_2$, then for any 
separable state the inequality
\begin{equation}
  \var(u) + \var(v) \geq \alpha^2 \abs{\mean{Z_1}} + \beta^2 \abs{\mean{Z_2}}
  \label{eq:entcriterion}
\end{equation}
holds. Choosing $\alpha=\beta=1$, $X_j=J^x_j$ and $Y_j=(-1)^{j+1}J^y_j$,
where $J^\nu_j$ is the $\nu$-component ($\nu=x,y,z$) of the total nuclear angular momentum $\bm{J}_j \equiv \sum_i \bm{I}_{ij}$, 
we conclude that if $\Delta_\mathrm{EPR}<1$, where
\begin{equation}
  \Delta_\mathrm{EPR} \equiv 
  \frac{\var(J^x_1+J^x_2) + \var(J^y_1+J^y_2)}{|\mean{J^z_1}| + |\mean{J^z_2}|} 
\end{equation}
is the so-called EPR uncertainty, then there is entanglement between the two nuclear ensembles. Notice that this is a sufficient condition for the state to be entangled, but it is not necessary, i.e., there are entangled states that cannot be detected by this criterion. Experimentally, this criterion requires the measurement of the nuclear polarizations of each ensemble. 
This may be done using NMR techniques \cite{xue2011}, although the small 
relative size of even the largest ensembles in typical QDs makes it 
challenging to detect the nuclear signal of only the relevant nuclei---those that couple to 
the quantum dot electrons. On the other hand, for small ensembles ($N_j<10$) one may use 
other techniques \cite{jiang2009,london2013} that allow single nucleus addressing

Instead, the most natural way to measure observables of the nuclear spin 
ensemble in a QD is via the interaction with the electron spin, i.e., by 
measuring the Overhauser field \cite{chekhovich2013}. 
This allows to magnify the small nuclear signal and ensures that only spins actually coupled to the electron contribute. This technique, however, does not give access to the total nuclear spin operators if the couplings are inhomogeneous. But the principle described above, Eq.~\eqref{eq:entcriterion},
is more general and also applies to the collective operators $\bm{A}_j$; 
the variances of non-local operators involving these then have to be 
compared to expectation values of $[A^x_j,A^y_j]=iA^{(z,2)}$, 
$A^{(z,2)}_j\equiv\sum_ig_{ij}^2 I^z_{ij}$. Thus, we define a new quantity
\begin{equation}
  \Delta \equiv \frac{\var(A^x_1+A^x_2) + \var(A^y_1+A^y_2)}
  {|\mean{A^{(z,2)}_1}| + |\mean{A^{(z,2)}_2}|} \,,
\end{equation}
such that $\Delta<1$ implies that the two ensembles are entangled. 
To avoid confusion, we will refer to the expectation value $\mean{J^z_j}/(N_j I)$ simply as the polarization of the $j$th ensemble, while we will refer to $2\mean{A^{(z,2)}_j}$ as the generalized polarization (the range of both quantities is [-1,1]).
Note that for an homogeneous system with identical ensembles ($N_1=N_2$) both $\Delta_\mathrm{EPR}$ and $\Delta$ are equal.

One way to measure $A^z_j$ in a gate-defined single QD is to measure electron spin resonance in an external magnetic field $B_z$, e.g., by determining the resonance frequency at which spin blockade of transport through the QD is lifted \cite{klauser2006}. By rotating the nuclear spins using a suitable NMR pulse the same method can be used to 
measure $A^{x,y}_j$ (and their variances). However, there seems to be no direct access to $A^{(z,2)}_j$. One possibility is to exploit the Jaynes-Cummings-like coupling, $H^\perp_j$, of the electron spin to the nuclear ensemble. Neglecting $H^z_j$ for the moment, the probability for an electron spin prepared in the state ``down'' to flip after a short time $\tau<1/g_j$ is, to leading order in $\tau$, approximately given by $\tau^2g^2_j\mean{A^+_jA^-_j}/4$, and similarly the one for the flip of an initially spin-``up'' electron is $\tau^2g^2_j\mean{A^-_jA^+_j}/4$. The difference of the two gives $\tau^2g^2\mean{[A^+_j,A^-_j]}/4=\tau^2g^2\mean{A^{(z,2)}_j}/2$. 
The presence of $H^z_j$ in the hyperfine Hamiltonian leads to corrections of third and higher order, 
$-i\tau^3g^3\mean{[A^z_j,\{A^+_j,A^-_j\}]}/16$, which could be made even smaller if we add an external magnetic field (pointing in the z-axis) so as to compensate the Overhauser field $\langle A^z_j\rangle + g\mu_B B_\mathrm{ext}=0$. In that case, the correction would be given by $-i\tau^3g^3\mean{[\delta A^z_j,\{A^+_j,A^-_j\}]}/16$ ($\delta A^z_j\equiv A^z_j - \mean{A^z_j}$), which is small for the highly polarized states we are considering.

Besides the criteria discussed here, there are many other ones one could employ to detect bipartite entanglement experimentally, each of which comes with its own challenges in how to implement the measurements and to which states the criterion is most sensitive. For example, for two single-spin-1/2 
ensembles one could detect entanglement measuring the singlet occupation 
\cite{thekkadath2016}. Another more general option is to employ criteria based
on the covariance matrix \cite{gittsovich2008,tripathi2020}.

% In some cases, specially when the number of nuclei is small, the aforementioned 
% criteria may not be able to detect entanglement, yet the steady state is entangled,
% as can be shown theoretically, for example by the PPT criterion (see Supp.\ Material). 
% Experimentally, one can use other entanglement witnesses, such as the one based on 
% the singlet fraction \cite{thekkadath2016}, or different criteria based on the 
% covariance matrix \cite{gittsovich2008,tripathi2020}\dots

\section{Homogeneous case \label{sec:homogeneous}}

\begin{figure*}
  \includegraphics[width=\textwidth]{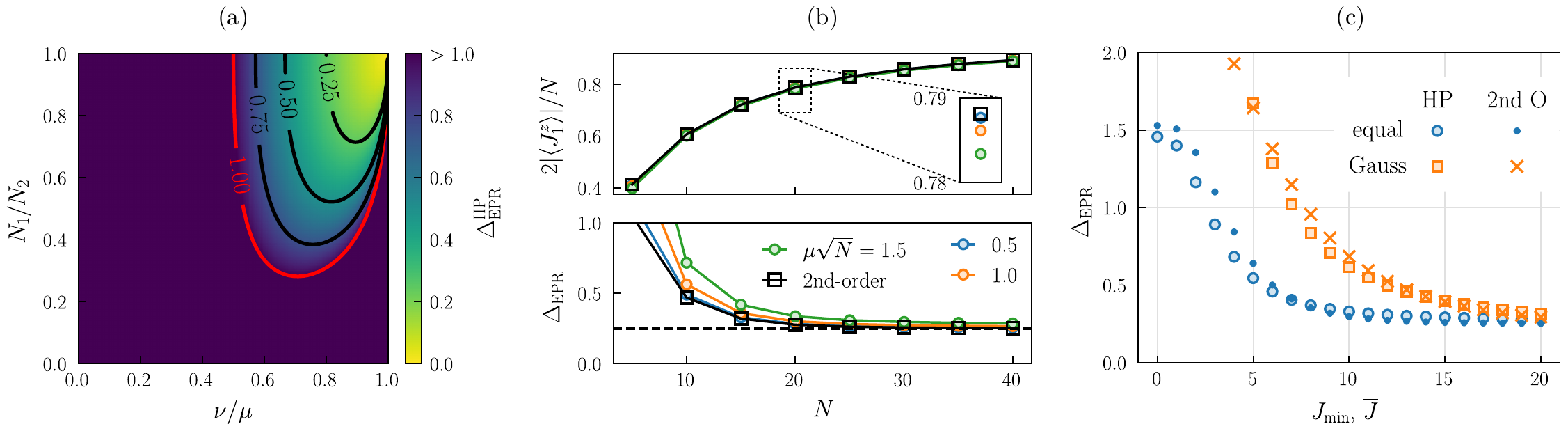}
  \caption{(a) Contour plot of $\EPR^\mathrm{HP}$, 
  Eq.~\eqref{eq:HPDeltaEPR}, for the subspace with maximal total angular
  momenta $J_j=N_jI$ ($r_j=1$). 
  (b) Characterization of the steady state in the subspace with maximum 
  total angular momenta as a function of the number of
  nuclei, for two identical ($N_1=N_2\equiv N$), homogeneous ensembles of spin-1/2 nuclei. We compare the exact dynamics (circles) with the result given by the second-order master equation (squares), which does not depend on the individual values of $\mu$ and $\nu$ but only on their ratio. For all the curves plotted this ratio is the same, $\nu/\mu = 0.8$. The dashed horizontal line marks the value obtained by the Holstein-Primakoff approximation in the limit $J_j\to\infty$, $\EPR^\mathrm{HP}$.
  (c) Steady-state $\EPR$  for an initial mixed state: blue circles and dots correspond to an initial equal-weight mixture of all states with $J_1,J_2 \geq J_\mathrm{min}$, while orange squares and crosses correspond to an initial state with a Gaussian probability distribution, $p_j(J)\propto \exp[-(J-\overline{J})^2/(2w^2)]$, of the total angular momenta with mean $\overline J$ and variance $w^2=3$. The system is the same as in panel (b), with $N = 40$ spins. The ratio $\nu/\mu = 0.8$ is the same for all the points. The results have been obtained using the Holstein-Primakoff approximation and the 2nd-order master equation.
  \label{fig:DeltaEPR_and_ss}}
\end{figure*}

To demonstrate that this protocol indeed produces entanglement, we begin analyzing the dynamics in the homogeneous case, i.e., when all nuclear spins belonging to the same ensemble couple with the same strength to the electronic spin ($g_{ij} = 1/\sqrt{2IN_j}$, for all $i=1,\dots,N_j$). 
In that case, the collective nuclear spin operators form a spin algebra 
$\bm{A}_j \propto \bm{J}_j$, and the total angular momentum of each 
ensemble $J_j\in \{N_j/2, N_j/2 - 1, \dots, (N_j \mod 2)/2\}$ 
% $\in \{0,1,\dots,N_j/2\}$ 
% ($I_j \in \{1/2,3/2,\dots,N_j/2\}$ if $N_j$ is odd) 
is conserved throughout evolution. Then, the density matrix can be split in blocks corresponding to different sectors (subspaces) with different values of $(J_1, J_2)$ that evolve independently, and the steady state has no correlations between these blocks. 

The last two dissipators of Eq.~\eqref{eq:mastereq} drive the spin ensembles towards an entangled
steady state, while the first two dissipators locally depolarize the nuclear spin ensembles and are detrimental for 
entanglement generation. If evolution were dictated just by the second line of Eq.~\eqref{eq:mastereq} (limit $k\to\infty$), then the steady state in the subspaces with equal total angular momenta $J_1=J_2=J$ would be a two-mode spin squeezed state 
\cite{ma2011}, given by the pure (unnormalized) state
\begin{equation}
  \ket{\Psi_\mathrm{TMS}} = \sum_{n=0}^{2J}
  \left(-\frac{\nu}{\mu}\right)^n \ket{J,J-n}_1 \ket{J,n-J}_2\,,
  \label{eq:psi_ideal}
\end{equation}
written here in terms of Dicke states $\ket{J, M}_j$ \cite{arecchi1972}, which are eigenstates of $\bm{J}_j^2$ and $J^z_j$ with eigenvalues $J(J + 1)$ and $M$ respectively. In this state, for sufficiently different $\mu$ and $\nu$ the two nuclear ensembles are highly polarized in opposite directions, and the EPR uncertainty turns out to be independent of $J$, 
$\Delta_\mathrm{EPR}=(\mu-\nu)^2/|\mu^2-\nu^2|$, hence 
$\Delta_\mathrm{EPR}<1$ for $\mu \nu> 0$, so the steady state is entangled. 
For unequal total angular momenta $J_1\neq J_2$, the steady state 
corresponding to the dissipators in the second line of Eq.~\eqref{eq:mastereq} 
is not pure, although still entangled \cite{schuetz2014}. For non-maximal $J_1,J_2$ the steady state is not unique, 
but the (spin permutation) quantum numbers associated with this non-uniqueness \cite{arecchi1972} are conserved by the homogeneous dynamics and do not affect the EPR uncertainty, so they can be ignored.

By increasing the number of steps per cycle (make $k$ larger), we are 
able to bring the actual steady state of Eq.~\eqref{eq:mastereq} closer to $\ket{\Psi_\mathrm{TMS}}$ (see Supp.\ Material). There is, however, a physical limitation on how small the dwell times can be made while still retaining control of the process. For example, in Ref.~\cite{takada2019}, the injection/ejection pulses take $\sim \SI{90}{ps}$, and the time-jitter is quoted as $\SI{6.6}{ps}$.
Nonetheless, as we show in the following, the protocol also works for small values of $k$ (even for $k=1$). 
% \sout{For the rest of the article we will focus on the protocol with $k=1$, 
% i.e., a four-steps protocol.}

As extracted from the analysis of the dynamics in the limit $k\to \infty$, the two nuclear 
ensembles become polarized in opposite directions. This is also true when 
considering a finite value of $k$. In this case,
we can obtain an analytical estimate of $\Delta_\mathrm{EPR}$, valid in the limit of large $J_j$, bosonizing the excitations of each ensemble over the fully polarized 
states $\ket{J_1,J_1}_1$ and $\ket{J_2,-J_2}_2$ 
% (valid for large $J_j$)
% $N_j$ limit bosonizing the excitations over the fully polarized
% state\Comment{GG: it would be enought to consider large $I_1,I_2$, no
%   need for full polarization} 
using the Holstein-Primakoff (HP) transformation \cite{holstein1940}, 
which maps spin operators $\bm{J}_j$ to bosonic operators 
$a_j$, $a_j^{\dagger}$ as
\begin{gather}
  J^z_1 = J_1 - a^\dagger_1 a_1 \,, \ 
  J^+_1 = \sqrt{2J_1 - a^\dagger_1 a_1}\, a_1 \,, 
  \\
  J^z_2 = a^\dag_2 a_2 - J_2 \,, \
  J^-_2 = \sqrt{2J_2 - a^\dag_2 a_2}\, a_2 \,.
\end{gather}
Note that the role of the ladder operators is reversed in QD2. If 
$\mean{a^\dag_j a_j} \ll J_j$, we can approximate 
\begin{equation}
  A^+_1 = \frac{J^+_1}{\sqrt{2N_1I}} \approx \sqrt{r_1} a_1 \,, \
  A^-_2 = \frac{J^-_2}{\sqrt{2N_2I}} \approx \sqrt{r_2} a_2 \,,
\end{equation}
and similarly $A^-_1\approx\sqrt{r_1}a^\dag_1$ and
$A^+_2\approx\sqrt{r_2}a^\dag_2$, where $r_j=J_j/(N_jI)$ is the
ratio between the total spin and the maximum possible total spin for
the $j$th ensemble. 
Rewriting Eq.~\eqref{eq:mastereq} using this transformation we 
obtain a new master equation whose steady state can be found 
analytically (see Supp.\ Material). For $k=1$, the EPR uncertainty in the steady state reads
\begin{equation}
  \Delta_\mathrm{EPR}^\mathrm{HP} = \frac{\mu^2}{4\kappa} - 1
  - \frac{\sqrt{J_1J_2}}{(J_1+J_2)}
  \frac{\sqrt{r_1r_2}}{(r_1+r_2)} \frac{\mu\nu}{\kappa}\,,
  \label{eq:HPDeltaEPR}
\end{equation}
where $\kappa\equiv(\mu\nu - \nu^2)/4>0$.
% Note that any two non-negative numbers, $x$ and $y$, satisfy 
% $(x + y)/2 \geq \sqrt{xy}$, with equality if and only if $x = y$. Thus,
Considering the relation between the arithmetic and geometric means,
we can readily see that $\Delta_\mathrm{EPR}^\mathrm{HP}$ is smaller the 
more similar the total spins $J_1$ and $J_2$ are, and also the more similar the ratios $r_1$ and $r_2$ are. In Fig.~\ref{fig:DeltaEPR_and_ss}(a) we plot this quantity for the subspace with maximal total angular momenta ($J_j=N_jI$), showing the values of $N_1/N_2$ and $\nu/\mu$ for which we expect to detect entanglement in the steady state. 

In addition, this bosonization approach allows us to compute the dynamics, provided the conditions $\mean{a^\dag_j a_j} \ll J_j$ are satisfied at 
all times. This is indeed the case for sufficiently small $\nu/\mu$ and for the initial fully polarized state 
$\ket{\Psi_\mathrm{FP}} = \ket{\Uparrow}_1 \ket{\Downarrow}_2$, $\rho_0 = 
\ket{\Psi_\mathrm{FP}} \bra{\Psi_\mathrm{FP}}$ ($\ket{\Uparrow}_j$, 
$\ket{\Downarrow}_j$ denote the states in which all nuclei in QD$j$ are 
polarized ``up'' or ``down'' respectively). The resulting dynamics 
approaches the steady state polarization exponentially with rate 
$\kappa r_j$ in each QD (see Supp.\ Material). If both ensembles are initially
highly polarized ($r_1, r_2\simeq 1$), we can take the associated time 
$T_\mathrm{stab}^\mathrm{hom} \equiv 1/(2\kappa)$ as an estimate of the timescale in which the system 
reaches the steady state.  

In Fig.~\ref{fig:DeltaEPR_and_ss}(b) we compare the steady-state polarizations 
and $\Delta_\mathrm{EPR}$ obtained by means of the exact Eq.~\eqref{eq:exactevol} 
with the ones obtained using the second-order master equation, 
Eq.~\eqref{eq:mastereq}, for a system consisting of two identical, homogeneous ensembles (initially fully polarized). While the polarizations do not present visible differences, there is some discrepancy between both in terms of entanglement, which is reduced by making the interaction times smaller. As we can see, $\EPR$ decreases monotonically as we increase $N$ ($J_j$), and the approximation $\Delta^\mathrm{HP}_\mathrm{EPR}$ 
agrees well with the result of the master equation already for 
$N \gtrsim 25$. Thus, in order to obtain the maximum amount of entanglement it is preferable that initially the subspace with largest occupation be the one with maximum total angular momenta 
$(J_1, J_2) = (N_1I, N_2I)$. This requirement on the initial state could 
be enforced polarizing completely the two nuclear ensembles, a task for 
which many protocols have been devised
\cite[e.g.][]{christ2007,GK+Lukin13,LL+Platero13,EcBa14,KW+Lesanovsky15} 
and towards which significant experimental progress has been reported
\cite{Nichol2015,MM+Marcus16,PH+Ludwig12,CU+Skolnick17}. Some limitations 
to dynamical nuclear polarization have also been predicted \cite{HKBR14}.
% Also, we show how the approximation $\Delta^\mathrm{HP}_\mathrm{EPR}$ 
% agrees well with the result of the master equation already for 
% $N \gtrsim 25$.

We remark that the protocol also works for non-maximal $J_j$ and even mixed initial states. For example, we may consider an initial state of the form $\rho_0 = \rho_\mathrm{QD1}\otimes\rho_\mathrm{QD2}$, with $\rho_{\mathrm{QD}j} = \sum_J p_j(J) \rho_j(J)$. Here, $\rho_j(J)$ is some normalized density matrix that belongs to the subspace of states of the $j$th nuclear ensemble with total angular momentum $J$, and $p_j(J)$ is some discrete probability distribution. Since the steady state within each fixed angular momenta subspace is unique (up to variations of the permutation quantum numbers), the steady-state value of $\Delta_\mathrm{EPR}$ only depends on the initial distributions $p_j(J)$. In Fig.~\ref{fig:DeltaEPR_and_ss}(c) we show the resulting steady-state $\EPR$ for an initial Gaussian distribution
% \begin{equation}
%     p_j(J) = \mathcal{N} \exp\left(-\frac{\left(J - \overline{J}\right)^2}{2w^2}\right)\,,
% \end{equation}
and also for an equal-weight mixture of all states with $J_1,J_2\geq J_\mathrm{min}$, demonstrating that maximal polarization of the nuclear ensembles is not strictly required for the protocol to work (see Supp.\ Material for further details).

\section{Inhomogeneous case \label{sec:inhom}}

\begin{figure}[H]
  \centering
  \vspace*{2em}
  \includegraphics[width=0.9\linewidth]{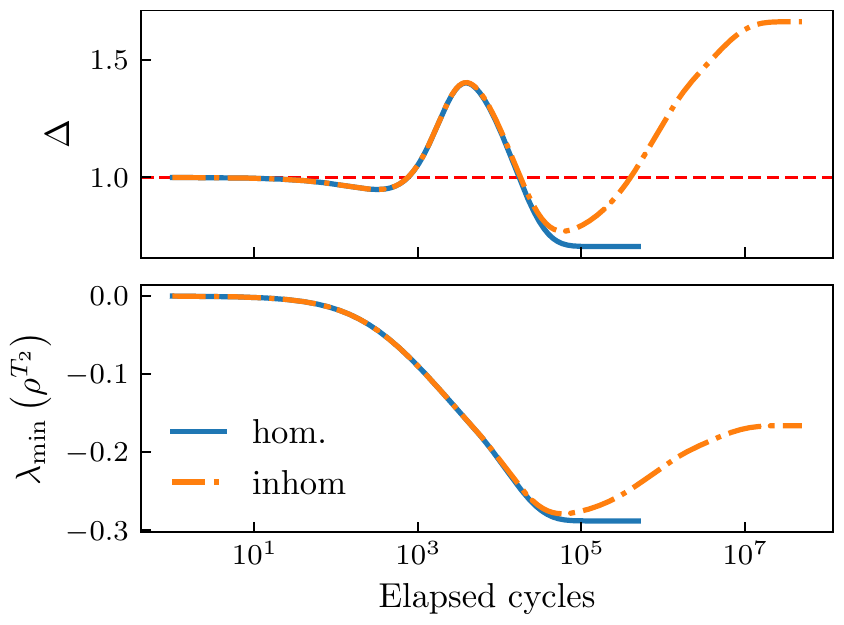}
  \caption{Simulation of a small system with $N_1=N_2=2$ nuclei in each
  ensemble computed using the exact dynamics, Eq.~\eqref{eq:exactevol}. The initial 
  state is $\ket{\Psi_\mathrm{FP}}$. The rest of 
  parameters are $\mu=0.1/\sqrt{N_1}$, $\nu=0.8\mu$, $k=5$. 
  Comparison between the homogeneous case, $g_{i1} = g_{j2} = 1/\sqrt{2}$ (blue),
  and a case with a small inhomogeneity, $g_{11} \simeq 0.714, g_{21} \simeq 0.7 , g_{12} \simeq 0.725, g_{22} \simeq 0.689$ (orange). Even though $\Delta>1$ for the inhomogeneous case,
  the steady state is entangled, as it is shown in the bottom panel, where we plot the lowest eigenvalue of partial transpose nuclear density matrix with respect to the nuclei in QD2, $\rho^{T_2}$. Note how the small difference between
  the couplings in the two cases considered only affects the dynamics at long
  times.\label{fig:simulationsmall}}
\end{figure}

As explained in the beginning of Section~\ref{sec:homogeneous}, the homogeneous case is numerically simpler because of the conservation of total angular momentum. However, in the inhomogeneous case, the total angular momentum of each nuclear ensemble is not conserved and we have to use the full Hilbert space in order to compute the dynamics. 
% In this section we address the dynamics induced by the protocol when the nuclear spins of each ensemble couple inhomogeneously to the electronic spins. The main complication we face in this case is that the total spin of each ensemble is not conserved, making it necessary to consider the full system's Hilbert when using Eqs.~\eqref{eq:exactevol} or \eqref{eq:mastereq} to compute the dynamics. 
This limits the direct application of Eqs.~\eqref{eq:exactevol} or \eqref{eq:mastereq} to very small systems. Henceforth, we will consider spin-1/2 nuclei,
$I^\alpha_{ij}=\sigma^\alpha_{ij}/2$ ($\alpha=x,y,z$). 
In Fig.~\ref{fig:simulationsmall} we show the expected dynamics for a 
system with only a few spins in each QD, considering both homogeneous
and inhomogeneous couplings. For inhomogeneous couplings the criterion 
based on $\Delta$ fails to detect entanglement, yet the steady state is 
entangled as can be demonstrated by the PPT criterion \cite{peres1996,horodecki1996,guhne2009}. 
% \red{Experimentally this could be demonstrated measuring the covariance matrix 
% (I've tested Eq.~(46) of \cite{gittsovich2008})}

For larger ensembles, we concentrate instead on the quantities required to compute $\Delta$, i.e., $\mean{A^{(z,2)}_{1,2}}$ and 
$\var(A^{x,y}_1 + A^{x,y}_2)$. They depend on the elements of the covariance matrix $\gamma^{jj'}_{ii'}\equiv\mean{\s{+}{ij}\s{-}{i'j'}}$ and lower-order terms like $\langle\s{x}{ij}\rangle$, whose equations of motion (EoMs) can be derived from Eq.~\eqref{eq:mastereq}
as (see Supp.\ Material)
\begin{widetext}
\begin{multline}
  \Delta\gamma^{jj'}_{ii'} = \left(\kappa - \frac{\mu^2}{4}\right)
   \left(g_{ij}^2 + g_{i'j'}^2\right)\gamma^{jj'}_{ii'}
  + \left[\mu^2\delta_{jj'} + \mu\nu\left(1 - \delta_{jj'}\right)
  - 8\kappa \delta_{j2}\delta_{j'2}\right] \frac{g_{ij}g_{i'j'}}{4}
  \mean{\s{z}{ij}\s{z}{i'j'}} \\
  + \kappa (-1)^j g_{ij} \sum_{k\neq i} g_{kj}
  \mean{\s{+}{kj}\s{z}{ij}\s{-}{i'j'}} 
  + \kappa (-1)^{j'} g_{i'j'} \sum_{k\neq i'} g_{kj'}
  \mean{\s{+}{ij}\s{z}{i'j'}\s{-}{kj'}}\,,
  \label{eq:EoMgamma}
\end{multline}
\begin{multline}
  \Delta \lambda^{jj'}_{ii'}= \left[
    4\kappa \left(\delta_{j2}g^2_{ij}+\delta_{j'2}g^2_{i'j'}\right)
    - \frac{\mu^2}{2}\left(g^2_{ij} + g^2_{i'j'}\right)\right]
  \lambda^{jj'}_{ii'}
  +\left[\mu^2\delta_{jj'}+\mu\nu\left(1 - \delta_{jj'}\right)
  - 8\kappa \delta_{j2}\delta_{j'2}\right]g_{ij}g_{i'j'}
  2\gamma^{jj'}_{ii'}
  \\
  -2\kappa(-1)^j\sum_k g_{kj}g_{ij}
  \left(\mean{\s{+}{ij}\s{z}{i'j'}\s{-}{kj}}
    +\mean{\s{+}{kj}\s{z}{i'j'}\s{-}{ij}}\right) 
  - 2\kappa (-1)^{j'}\sum_k g_{kj'}g_{i'j'}
  \left(\mean{\s{+}{i'j'}\s{z}{ij}\s{-}{kj'}}
    +\mean{\s{+}{kj'}\s{z}{ij}\s{-}{i'j'}}\right) ,
  \label{eq:EoMlambda}
\end{multline}
\end{widetext}
As can be seen in these equations, the elements of the covariance matrix depend on $zz$-correlations, $\lambda^{jj'}_{ii'}\equiv\mean{\s{z}{ij}\s{z}{i'j'}}$
for $(i,j)\neq (i',j')$, and higher-order correlations. 
Due to the spin-operator commutation relations, and since the Lindblad operators depend on linear combinations of $\sigma_{ij}^\pm$ only, the only higher-order terms occurring in the EoM for $\gamma^{ii'}_{jj'}$ and $\lambda^{ii'}_{jj'}$ are of the form $\langle\sigma^+_i \sigma^z_j \sigma^-_k\rangle$. %This holds for arbitrary spin nuclei \red{check!}.
In order to obtain a tractable system of equations, it is necessary to 
introduce some factorization assumption which effectively approximates these higher-order correlations in terms of lower-order ones. There are several approximation schemes one can use for this purpose \cite{christ2007}. 
Among all the approximations we have investigated (see Supp.\ Material), the one which best approximates the dynamics 
%given by the second-order master equation 
is
\begin{equation}
  \mean{\s{+}{ij}\s{z}{kl}\s{-}{mn}} \approx 
  \left(2\gamma^{ll}_{kk} - 1\right)\gamma^{jn}_{im} 
  + 2 (-1)^l \gamma^{jl}_{ik} \gamma^{ln}_{km}\,,
\end{equation}
for $(i, j) \neq (k, l) \neq (m, n)$ and $(m, n) \neq (i, j)$.

\begin{figure}
  \centering
  \includegraphics{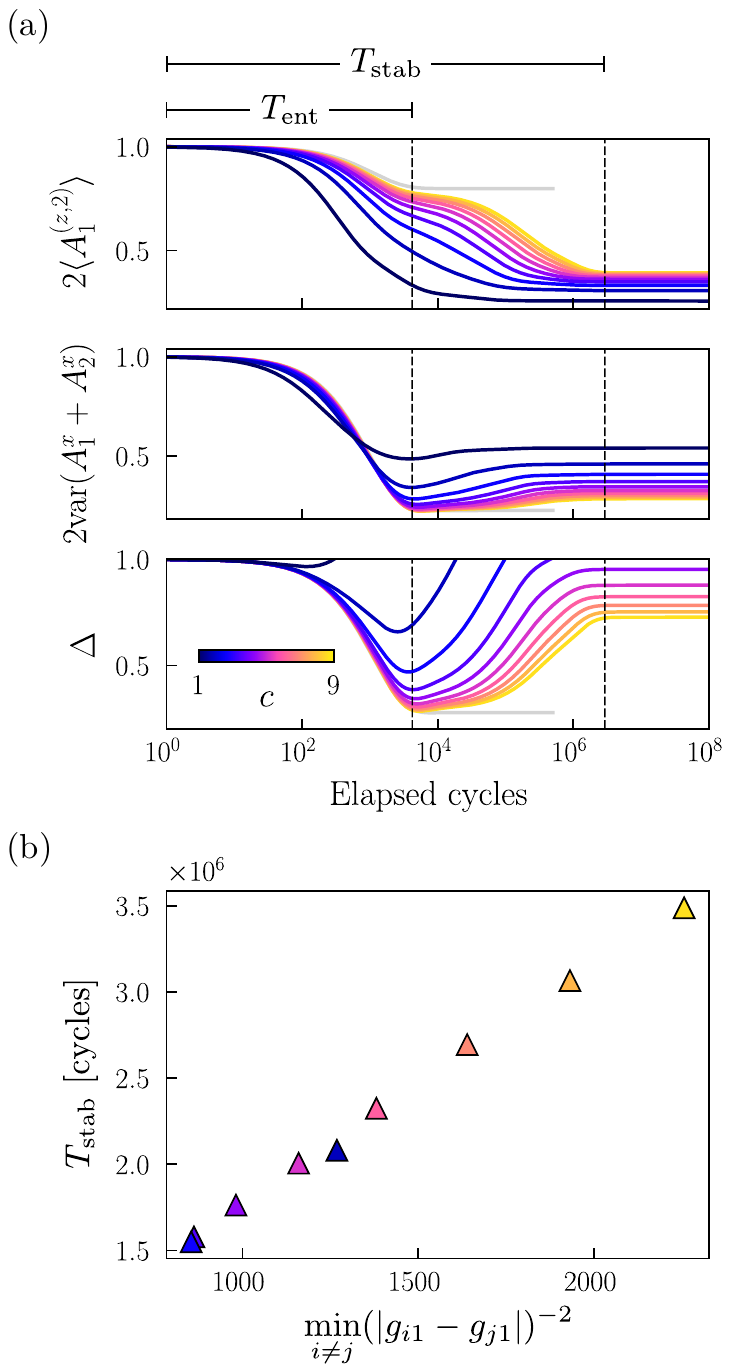}
  \caption{(a) Dynamics of a shell model with different inhomogeneity 
  degrees (color-coded). The couplings follow a Gaussian function $g_{ij}
  \propto \exp(-\norm{\bm{r}_{ij}}^2/c)$, where $\{\bm{r}_{ij}\}$ 
  denote the positions of the nuclei, which are assumed to be forming a 2D 
  square lattice, $\bm{r}_{ij}\in\mathbb{Z}^2$. We only consider the nuclei within a given distance from the origin $\norm{\bm{r}_{ij}}<r^\mathrm{max}_j$,
  $r^\mathrm{max}_1 = 2.5$ and $r^\mathrm{max}_2 = 3$, resulting in a total of $N_1=21$ and $N_2=25$ nuclei in each ensemble. The inhomogeneity in each QD is determined by the ratio $c/r^\mathrm{max}_j$, such that larger ratios correspond to more homogeneous systems. 
  Parameters: $\mu=0.5/\sqrt{N_1}$, 
%   (this implies a different value of the
%   dwell times for different $\{g_{ij}\}$, but makes the order of $\Delta\gamma, 
%   \Delta\lambda\sim \mu^2$ the same for all cases studied)
  $\nu=0.8\mu$, $k=1$. The initial state is $\ket{\Psi_\mathrm{FP}}$.
  Dashed vertical lines roughly indicate the two different timescales present in the dynamics.
  Grey curves show the dynamics for an equivalent homogeneous system.
  (b) Stabilization times for the generalized polarization in QD1 vs the inverse squared smallest non-zero difference between the couplings in QD1. For very inhomogeneous systems the points obtained (not shown) do not follow this linear dependence. The color code used to denote different inhomogeneity degrees is the same in both panels. \label{fig:evolshells}}
\end{figure}
            
In Fig.~\ref{fig:evolshells}(a) we employ these EoMs to compute the dynamics of a large inhomogeneous system (larger than the one shown in 
Fig.~\ref{fig:simulationsmall}, but still much smaller than realized in  
experiments with typical GaAs or Si QDs). The results show that for not too inhomogeneous couplings the steady state is entangled, and that this entanglement can be detected experimentally via $\Delta$. Furthermore, for all cases considered $\Delta<1$ after some initial cycles, albeit at later times $\Delta$ takes on larger values and deviates away from this minimum, reaching $\Delta>1$ for the more inhomogeneous cases. 
For moderately inhomogeneous systems there is a clear distinction between two timescales. On the one hand, there is the timescale in which the system reaches the minimum $\Delta$, which roughly corresponds to the stabilization time of the equivalent homogeneous system, $T_\mathrm{ent}\sim (2\kappa)^{-1}$. 
On the other hand, there is the timescale in which the system reaches the steady state, $T_\mathrm{stab}$, which is now much longer than in the homogeneous case. Also, estimating this timescale is harder in the inhomogeneous case, as the dynamics do not follow a simple exponential decay. The dynamics of the generalized polarization of each ensemble, $\mean{A^{(z,2)}_j}$, is perhaps the simplest to analyze, as it depends just on the coupling constants within the given ensemble. Numerically, we can show that the stabilization time scales with the smallest difference of the coupling constants of the corresponding ensemble as $T_\mathrm{stab}\sim \left[\mu\min(\abs{g_{ij}-g_{i'j}})\right]^{-2}$,
see Fig.~\ref{fig:evolshells}(b), where $T_\mathrm{stab}$ has been computed as the time in which
$\mean{A^{(z,2)}_j}_{T_\mathrm{stab}} - \mean{A^{(z,2)}_j}_\infty = 10^{-3}\mean{A^{(z,2)}_j}_\infty$.

Besides computing the dynamics, we can solve directly for the steady state, that is, solve the system of equations 
$\Delta\gamma^{jj'}_{ii'} = 0$ and $\Delta\lambda^{jj'}_{ii'} = 0$. Note 
that for the initial state $\rho_0 = \ket{\Psi_\mathrm{FP}} \bra{\Psi_\mathrm{FP}}$, many of the covariance matrix elements and $zz$-correlations are equal: 
$\gamma^{11}_{ii}=1$, $\gamma^{22}_{ii}=0$, $\gamma^{jj'}_{ii'}=0$,
for $(i,j)\neq(i',j')$; $\lambda^{jj}_{ii'}=1$, and $\lambda^{12}_{ij}=-1$.
We realize that for this initial condition, if two nuclei that belong to the same QD have the same coupling, say $g_{ij}=g_{i'j}$, then 
$\gamma^{jl}_{ik}=\gamma^{jl}_{i'k}$ and 
$\lambda^{jl}_{ik}=\lambda^{jl}_{i'k}$ for all $k\in\mathrm{QD}l$ throughout the entire time evolution. Therefore, we can reduce the number of variables considering just one representative for each possible value of the couplings. This is what we will refer to as ``shell model'', 
% the shells being each of the sets of nuclei with the same coupling
% constant.
the shells being the sets of nuclei within each QD with the same coupling
constant.

Numerically, we find for this initial state that correlations between 
nuclear spins in different shells of the same QD vanish in the steady state.
As a consequence, the steady-state polarization of each shell is independent of the others, and can be found analytically solving a simple quadratic equation (see Supp.\ Material). Furthermore, the steady-state polarization of each shell only depends on the values of $\mu$, $\nu$, and the number of nuclei in the shell, but not on the specific value of the coupling constants. 
% {\color{red} This holds as long as we can neglect all nuclear-spin-relaxation processes (dipolar, phonons, fluctuating magnetic fields, etc.), i.e., on short timescales compared to the nuclear spin lifetimes (\SI{100}{\micro\second} to \SI{1}{\second} depending on the setting)} 
% \Comment{GG: i don't understand why the number of nuclei
%   matters: it seems the ss polarization describes the balance between
%   the up- and down-flipping processes in the master eq.; since the
%   coupling constant (and a potential collective enhancement from
%   $N_j>1$) affect both processe equally, the ratio is determined by
%   $\mu$ and $\nu$ only. What am I missing?} 
On the other hand, the steady state value of the transverse spin variance depends continuously on the specific values of the coupling constants.

The condition that two nuclei have precisely the same coupling constant 
is somewhat exceptional in a real experiment, yet the steady state 
depends on this condition, as it determines the number of 
shells and the number of nuclei in each shell.
In any case, small variations in the couplings only affect the long-term 
dynamics, making it possible to have an entangled state at intervening 
times for almost any system with sufficiently homogeneous couplings, as Figs.~\ref{fig:simulationsmall} and \ref{fig:evolshells} suggest (see the Supp.\ Material for further calculations supporting this conclusion).

% \Comment{GG: can we quantify these timescales? If already for the rather % strong coupling for $25$ spins we need $10^4$ cycles, how much more
% would be need for $N=10^6$?} 

% \Comment{GG: having to go to $10^8$ cycles to reach the steady state seems 
% to reflect really tiny eigenvalues of the Liouvillian (or the CP map given 
% by eq.(2)): should't we check whether on that timscale terms we neglected 
% completely (such as dipolar coupling, non-contact terms or electron 
% decoherence) might be more important?}

\section{Experimental realizations \label{sec:experiment}}

\begin{table}[t]
  \caption{Typical time- and energy-scales for different possible experimental setups. We consider the case of GaAs QDs, and Si QDs with different concentrations of \isotope{Si}{29}. In all the cases shown the parameters of the protocol are the same: $N_1\simeq N_2\simeq N, \mu=1.5/\sqrt{2IN}, \nu=0.8\mu$ and $k=1$. Furthermore, we assume approximately homogeneous couplings $g_{ij}\simeq 1/\sqrt{2IN}$, $g_j\simeq A_H\sqrt{2I/N}$. The time per cycle (taking into account only the dwell times) is given by $\Delta t=2\mu(g_1^{-1} + g_2^{-1})$, whereas the time in which we expect to produce entanglement is  $T_\mathrm{ent}\sim (2\kappa)^{-1}$ cycles, which is of the order of $\sim 11 I N$ cycles. In the last column we show the nuclear dephasing time due to nuclear dipole-dipole interactions. \label{tab:experiment}}
  \begin{ruledtabular}
  \begin{tabular}{cccccc}
    Material & $N$ & $A_H$ [\si{\micro\electronvolt}] & $\Delta t$ [\si{ns}] & $T_\mathrm{ent}$ [\si{ms}] & $T_\mathrm{dd}$ [\si{ms}]\\[0.2em] \hline
    GaAs & \num{e5} & 91.2 & 0.014 & 0.024 & 0.1 \rule{0pt}{1em}\\
    % GaAs & \num{e5} & \sout{10}\textcolor{red}{100} & 0.4 & 0.2 & \\
    20\% Si & \num{2e4} & 0.86 & 5 & 0.5 & 1.7 \\
    5\% Si & \num{5e3} & 0.22 & 18 & 0.5 & 7 \\
    0.1\% Si & 100 & \num{4.3e-3} & 918 & 0.5 & 349
  \end{tabular}
  \end{ruledtabular}
\end{table}

The characteristic timescales of our protocol depend strongly on the materials used to make the device. Table \ref{tab:experiment} summarizes at a glance the typical energy- and time-scales for different experimental setups. One of the most common material for QD fabrication is GaAs. All isotopes of both Ga and As have nuclear spin
$I=3/2$ \cite{schliemann2003}. Since the dwell times are inversely proportional to $A_H$, the large value $A_H \sim \SI{91.2}{\micro\electronvolt}$ 
% $A_H\sim \text{\SIrange{10}{100}{\micro\electronvolt}}$ 
(average value considering the different nuclear isotopes and their natural abundance) \cite{CU+Skolnick17} in GaAs imply short dwell times in the order of a few to tens of picoseconds.
%$\tau_j \sim \text{\SIrange{0.01}{0.1}{\nano\second}}$ for typical protocol parameters. 

% For example, a value of $\mu = 1.5/\sqrt{N}$, 
% $\nu=0.8\mu$, $k=1$, results in dwell times on the order of 
% $\tau_j \sim \text{\SIrange{0.01}{0.1}{\nano\second}}$ (for an homogeneous
% system).
% \Comment{GG: as mentioned above, in \cite{takada2019} the
%   injection/ejection pulses take $\sim90$ps, the time-jitter is quoted
%   as 6.6ps so the numbers do not look too hard (except to reach large $k$).}
% \sout{Our protocol is therefore more suitable to Si.} 

Another well-known material for QD fabrication is Si. Si has two isotopes, but only one of them is magnetic, \isotope{Si}{29}, with a natural abundance of around $5\%$ %$4.7\%$ 
and nuclear spin $I=1/2$. Furthermore, the concentration of \isotope{Si}{29} in Si can be controlled by different methods of enrichment or purification. This allows great flexibility in the design of the device, with ensemble sizes ranging from a few nuclei to tens of thousands of them. The
smaller hyperfine constant $A_H\simeq \SI{4.3}{\micro\electronvolt}$ (for Si with a 100\% concentration of \isotope{Si}{29})  \cite{assali2011} results in dwell times that can range from a few nanoseconds to a few microseconds depending on the concentration of spinful nuclei.

% The average number of nuclei in a typical QD is of the order of $N\sim \num{5e3}$ and 
% $A_H\sim \SI{0.21}{\micro\electronvolt}$ \cite{assali2011}. The dwell times in this case 
% are $\tau_j \sim \SI{1}{\nano\second}$ for the same values of $\mu, \nu$ and $k$.
% % , which is experimentally accessible. 
% For such system $T=(2\kappa)^{-1}\sim 5.6 N$ cycles, which in the homogeneous
% case corresponds to $T\sim \SI{0.1}{\milli\second}$ (the time per cycle is always 
% $\Delta t= 4\mu\sqrt{N} A_H^{-1}$ regardless the value of $\nu$ and $k$).
% but for an homogeneous system each cycle is 
% $\Delta t= 4\mu\sqrt{N} A_H^{-1}$, so. 
% Another advantage of using Si is that it can be enriched or purified, so the 
% number of magnetic nuclei can be greatly varied. For example, for a purified 
% system with $N\sim \num{e2}$, $A_H\sim \SI{4.3e-3}{\micro\electronvolt}$ \cite{assali2011}, the dwell times and stabilization time are $\tau_j\sim \SI{100}{\nano\second}$ and
% $T\sim \SI{0.1}{\milli\second}$, for the same values of $\mu, \nu$ and $k$. 

% For a setup such as the one of Hensen et al.\ \cite{hensen2020}, where hiperfine 
% couplings were measured to be at least $\SI{100}{kHz}$ \red{(this refers to $A_H$ 
% or the individual couplings $g_j g_{ij}$?)} then, for a protocol such as the one shown 
% in Fig.~\ref{fig:simulationsmall}, with $N_1=N_2=2$, $\mu=0.1/\sqrt{N_1}$,
% $\nu=0.8\mu$, and $k=5$, we would have dwell times as small as 
% $\tau_j\sim \SI{10e-9}{s}$.

% nuclear spin decay/decoherence
It is important to take into account other effects we have neglected up to now, which lead to nuclear spin decay and decoherence. The $T_1$ and $T_2$ times of nuclear spins are long, but finite, and protocols that are slow compared to these times cannot be expected to be well-described by our model (and thus, likely, are not able to produce entanglement). In this regard, the most relevant processes that would hinder the generation of entanglement are nuclear dipole-dipole interactions. In GaAs nearest-neighbor dipole-dipole interactions are estimated to be $\sim \SI{e-5}{\micro\electronvolt}$ \cite{schliemann2003,GK+Lukin13} corresponding to $T_\mathrm{dd} \sim \SI{0.1}{\milli\second}$; in Si,
where the distance between nuclei is larger and they have weaker magnetic moment, nearest-neighbor dipole-dipole interactions can be estimated as $\sim\SI{4e-7}{\micro\electronvolt}$ (for \isotope{Si}{29} at concentrations larger than 12.5\%), corresponding to $T_\mathrm{dd} \sim \SI{1}{\milli\second}$.
Furthermore, reducing the concentration of \isotope{Si}{29} would decrease also the strength of these unwanted interactions, leading to longer coherence times for the nuclei.
% \sout{Furthermore, at high polarizations, we
% expect these values to overestimate the effect of dipole-dipole
% interactions since only the coupling between oppositely oriented spins
% degrades the state \cite{Schwager2010}}. 
In GaAs QDs, entanglement is generated after tens of microseconds, while one needs hundreds of microsends for Si devices. This result is almost independent of the concentration of \isotope{Si}{29} because $T_\mathrm{ent}\propto N/A_H$. According to these estimations, it seems that reaching the highly entangled transient state is indeed feasible in both GaAs and Si. 
% {\color{red} The value of $T_\mathrm{dd}\sim T_\mathrm{ent}$ for GaAs is perhaps too tight to reach the transient highly entangled state, although some entanglement may still be generated using shorter runs of the protocol. 
% For Si (at natural concentrations of \isotope{Si}{29} and below), by contrast, $T_\mathrm{dd}<T_\mathrm{ent}$, which suggests that reaching this highly entangled state is indeed feasible.}

Ferromagnetic leads or spin filters can be used for spin polarized electron injection \cite{hanson2004,cota2005,busl2010}. Concerning the coherent electron bus, great progress has been achieved towards this goal using surface acoustic waves in piezoelectric materials 
\cite{bauerle2018,takada2019,jadot2020}, and transport across long QD arrays \cite{baart2016,ban2019,mills2019,yoneda2020}. Although Si is not piezoelectric, the former bus could be implemented combining Si with other materials in a layered structure \cite{buyukkose2013}. 
One important imperfection that may arise in experimental implementations is the dephasing of the electron spin during transport from QD1 to QD2. If the transported spin would fully dephase, it would induce two independent jump operators $\mu A^+_1$ and $\nu A^+_2$ instead of the combination $\mu A^+_1 + \nu A^+_2$ (and similarly for $\nu A^-_1 + \mu A^-_2$). This could be modelled including a dephasing channel \cite[p.\ 384]{NielsenChuang} between the action of $e^{\tau_1\mathcal L_1}$ and $e^{\tau_2\mathcal L_2}$ in the first two steps of each cycle. The new master equation would be as Eq.~\eqref{eq:mastereq}, 
with the non-local dissipator terms $\mathcal D(\mu A^+_1 + \nu A^+_2)$ and $\mathcal D(\nu A^-_1 + \mu A^-_2)$ re-scaled by a factor $(1 - p)$ and 
additional terms: $p\left[\mathcal D(\mu A^+_1) + \mathcal D(\nu A^+_2) + \mathcal D(\nu A^-_1) + \mathcal D(\mu A^-_2)\right]/4$. Note that two of these new terms add up to the local polarizing terms that we considered already, whereas the other two constitute new local depolarizing terms. If the strength of these new terms is much smaller than that of the local polarizing terms, $p\mu^2 \ll (\mu - \nu)^2/k$, then the polarizing terms dominate and we expect the nuclear dynamics to be very similar to the one in the limit of perfect coherent transport. 
One could estimate the phase-flip probability from the ensemble-averaged spin dephasing time of the electron spin during transport, which is averaged due to its motion (motional narrowing). The numbers reported in Ref.~\cite{jadot2020} show that surface-acoustic-wave transport in GaAs itself adds little dephasing, and an estimate of the dephasing time due to the motionally averaged Overhauser field of the
nuclei in the channel suggests a negligible phase-flip probability $p\sim 1\%$ for a channel of length \SI{5}{\micro\meter}. Further details about the effect of electron-spin decoherence during transport are given in the Supplementary Material (sections \ref{sec:appendix_effectiveME}, \ref{sec:appendix_HolsteinPrimakoff}, and  Fig.~\ref{fig:EPRgeneral}), where we show that even for a phase-flip probability $p\sim 5\%$ it is possible to obtain an entangled steady state (for an homogeneous system) depending on the rest of system parameters and the choice of dwell times. 

% One could estimate the phase-flip probability as $p = 1 - e^{-\delta t/2T_2^*}$, where $T_2^*$ is the ensemble-averaged spin-spin dephasing time of the electrons in the material. However, this estimation would be too restrictive since the effective Overhauser field experienced by the electron spin during transport is averaged due to its motion (motional narrowing), which means that the actual dephasing time is much larger. The numbers reported in \cite{jadot2020} show that surface-acoustic-wave transport in GaAs itself adds little dephasing, and an estimate of the dephasing time due to the motionally averaged Overhauser field of the
% nuclei in the channel suggests a negligible phase-flip probability $p\sim 1\%$ for a channel of length \SI{5}{\micro\meter}.

% 

The dissipative nature of our protocol makes it inherently robust against
perturbations such as imperfect control of the dwell times 
\cite{benito2016}, and the preparation of the initial state. For the 
same reason, it will drive inhomogeneous nuclear ensembles towards an 
entangled state as long as the inhomogeneity is not too strong. In this 
sense, this type of protocol is expected to be more robust than coherent 
protocols such as the one presented in Ref.~\cite{andersen2012}.

\section{Conclusions \label{sec:conclusion}}

In this work we propose a protocol for the deterministic generation of
entanglement between two distant ensembles of nuclear spins, which are 
located around two QDs connected via a coherent electron bus. The protocol relies on the ability to inject electrons with definite polarization in the QDs and the coherent transfer of the electrons from one QD to the other. 
Controlling the time the electrons spend trapped in each QD, we engineer a bath that drives the system towards an entangled steady state. The dissipative nature of our protocol makes it inherently robust against imperfections in the dwell times and the electron transport. 

Using a bosonization technique we have fully characterized the steady state in the homogeneous case within the large total angular momentum limit.
% {\color{red}\sout{Furthermore we have analyzed the steady state in the inhomogeneous case using the EoMs for the covariance matrix, and showed that, while having an entangled steady state is not guaranteed generally, for sufficiently homogeneous systems it is possible to produce stable entanglement using short runs of our protocol.}}
%
Furthermore we have analyzed the dynamics in the inhomogeneous case using the EoMs for the covariance matrix. There, we have identified two different time scales characterizing the protocol: a faster one related to the collective hyperfine coupling on which the spins are driven into an entangled state, as evidenced by spin-squeezing of the collective nuclear spin operators, and a slower one related to the inhomogeneity, which leads to partial dephasing of the entanglement.

While the protocol requires the coherent transport of electron spins---a resource which also allows to distribute electron-spin entanglement---employing the nuclear spins makes an additional resource available for quantum information processing: in a hybrid approach, constantly generated nuclear-spin entanglement could be a resource accessed by the electron-spin quantum register at need. To take full advantage of the long nuclear coherence time, it would be appealing to store quantum information in the nuclear system, using the electron spins only to operate on the nuclei. Note that our protocol generalizes directly to the case of $n$ spin ensembles. This is readily seen in the homogeneous and bosonic limit, where $n$ linearly independent terms akin to the two non-local ones in the master equation Eq.~\eqref{eq:mastereq} will stabilize an $n$-mode squeezed vacuum state, which comprise, for example, Gaussian cluster states that allow universal measurement-based quantum computation \cite{MLG+06}. Alternatively, pairs of entangled spin ensembles can be merged into multipartite entangled states by performing, for example, a measurement of a joint collective spin component of two of them; combined with full unitary control over the nuclear spins (as conceivable for few or single nuclear spins \cite{hensen2020}) this also allows the preparation of computationally universal \cite{RaBr01} multipartite entangled states.
% \appendix
% \section{}

\begin{acknowledgments}
%   The authors acknowledge long and fruitful discussions with Martin Schuetz, which helped initiate this project. 
%   M. Schuetz participated in this work prior to joining AWS\@.
  The authors thank J. I. Cirac for useful discussions at the beginning of the project.
  M. Benito acknowledges support by the Georg H. Endress foundation. M. Bello and G. Platero acknowledge support from the Spanish Ministry of Science and Innovation through grant MAT2017-86717-P\@.
  M. Bello acknowledges support from the ERC Advanced Grant QUENOCOBA (GA No. 742102).
  G. Giedke acknowledges support from the Spanish Ministry of Science and Innovation through grant FIS2017-83780-P and from the European Union (EU) through Horizon 2020 (FET-Open project SPRING Grant no. 863098).
\end{acknowledgments}

\bibliography{nuclei.bib}

%%%%%%%%%%%%%%%%%%%% Supplemental Materials %%%%%%%%%%%%%%%%%%%% 

% to reference equations, figs, etc. in other files
% \usepackage{xr}
% \externaldocument{filename}

\clearpage
\onecolumngrid
\begin{center}
\textbf{\large Supplemental Material}
\end{center}
% reset the counters
\setcounter{section}{0}
\setcounter{equation}{0}
\setcounter{figure}{0}
\setcounter{table}{0}
\setcounter{page}{1}
% \makeatletter
% reset equation and figure counters after each section
\counterwithin*{equation}{section} 
\counterwithin*{figure}{section} 
\counterwithin*{table}{section}
\renewcommand\thesection{\Alph{section}}
\renewcommand\thesubsection{\thesection.\arabic{subsection}}
\renewcommand{\theequation}{\thesection\arabic{equation}}
\renewcommand{\thefigure}{\thesection\arabic{figure}}
\renewcommand{\thetable}{\thesection\arabic{table}}
% \renewcommand{\bibnumfmt}[1]{[S#1]}
% \renewcommand{\citenumfont}[1]{S#1}

% \appendix
In this Supplemental Material we provide some details on the derivation of 
the effective master equation, the Holstein-Primakoff approximation for 
homogeneous systems, the derivation and further analysis of the 
equations of motion (EoMs) for the covariance matrix elements, and the shell model.

\section{Effective master equation \label{sec:appendix_effectiveME}}

In this section we derive a general master equation including the effect of decoherence during electron transport and static errors in the dwell times arising from uncertainties in the value of the coupling constants $g_j$. Decoherence during electron transport can be modelled as a phase damping channel with phase-flip probability $p$. For an electronic density matrix $\sigma$, this quantum channel is given by $\mathcal{E}_p[\sigma] = (1 - p)\sigma + p\sigma_z\sigma\sigma_z$ \cite[p.\ 384]{NielsenChuang}. Imperfect knowledge of the coupling constants will result in dwell times that deviate from the ideal prescription shown in Table \ref{tab:protocol} of the main text. Instead, here we consider the evolution generated by the protocol specified in Table \ref{tab:protocol_reallistic}, which is more general than the one discussed in the main text and contains it as a particular case.

\begin{table}[!htbp]
  \caption{Electronic polarizations and dwell times for the steps that make 
  up a single cycle of the protocol, taking into account the possible decoherence during electron transport and the imperfect knowledge of the coupling constants.   \label{tab:protocol_reallistic}}
  \begin{ruledtabular}
  \begin{tabular}{ccccc}
    step(s) & $\ket{s_\mathrm{in}}$ & $g_1 \tau_1$ & & $g_2 \tau_2$ \\[0.2em] \hline
    $1$ & $\ket{\up}$ & $\mu_1$ & $\mathcal{E}_p$ & $\nu_2$ \rule{0pt}{1em}\\
    $2$ & $\ket{\down}$ & $\nu_1$ & $\mathcal{E}_p$ & $\mu_2$ \\
    $3, \dots, 2 + k$ & $\ket{\down}$ & $(\mu_1 -\nu_1)/k$ & -- & 0 \\
    $3 + k, \dots, 2 + 2k$ & $\ket{\up}$ & 0 & -- & $(\mu_2 - \nu_2)/k$ \\
  \end{tabular}
  \end{ruledtabular}
\end{table}

In the first two steps of the protocol, the nuclear density matrix evolves as
\[\rho_{n+1} = \tr_\mathrm{el}\left(e^{\tau_1\mathcal L_1}\left[
  \mathcal E_p \left[e^{\tau_2 \mathcal L_2}\left[\chi_n\right]\right]\right]\right)\,.\]
Expanding the exponentials up to second order, we find
\begin{equation}
  \begin{split}
    \rho_{n+1} & \simeq \tr_\mathrm{el} \Bigg(\mathcal E_p [\chi_n] 
    - i\tau_1 \mathcal E_p [[H_1,\chi_n]] - i\tau_2 [H_2,\mathcal E_p[\chi_n]] \\
    & \qquad \qquad -\frac{\tau_1^2}{2} \mathcal E_p[[H_1,[H_1,\chi_n]]] 
    - \frac{\tau_2^2}{2} [H_2,[H_2,\mathcal E_p[\chi_n]]] 
    -\tau_1\tau_2 [H_2,\mathcal E_p[[H_1,\chi_n]]] \Bigg) \,,
  \end{split}\label{eq:expansion}
\end{equation}
where $\tr_\mathrm{el}(\cdot)$ denotes the trace over the electron degrees of freedom.
Note that for the initial states considered, $\chi_n = \rho_n\otimes\ket{\up}\!\bra{\up}$
or $\chi_n = \rho_n\otimes\ket{\down}\!\bra{\down}$, $\mathcal E_p[\chi_n] =
\chi_n$. Also, for any total density matrix $\chi$, $\tr_\mathrm{el}(\mathcal
E_p[\chi]) = \tr_\mathrm{el}(\chi)$. Thus, only the last term of Eq.~\eqref{eq:expansion} is affected by decoherence. Specifically, we have
\begin{align}
  \tr_\mathrm{el} (\chi_n) & = \rho_n \,, 
  \\
  -i \tau_1 \tr_\mathrm{el} ([H_1, \chi_n]) 
  & = \mp i \frac{\tau_1 g_1}{2} [A^z_1, \rho_n] \,, 
  \\
  -i \tau_2 \tr_\mathrm{el} ([H_2, \chi_n]) 
  & = \mp i \frac{\tau_2 g_2}{2} [A^z_2, \rho_n] \,, 
  \\
  -\frac{\tau_1^2}{2} \tr_\mathrm{el} ([H_1, [H_1, \chi_n]]) 
  & = \frac{\tau_1^2 g_1^2}{4} \diss(A^z_1)[\rho_n] 
  + \frac{\tau_1^2 g_1^2}{4} \diss(A^\pm_1)[\rho_n] \,, 
  \\
  -\frac{\tau_2^2}{2} \tr_\mathrm{el} ([H_2, [H_2, \chi_n]]) 
  & = \frac{\tau_2^2 g_2^2}{4} \diss(A^z_2)[\rho_n] 
  + \frac{\tau_2^2 g_2^2}{4} \diss(A^\pm_2)[\rho_n] \,, 
  \\
  - \tau_1\tau_2 \tr_\mathrm{el} ([H_2,\mathcal{E}_p[ [H_1, \chi_n]]]) 
  & = -\frac{\tau_1 g_1 \tau_2 g_2}{4} [A^z_2, [A^z_1, \rho_n]]  
  - \frac{\tau_1 g_1 \tau_2 g_2(1-2p)}{4} 
  ([A^\mp_2, A^\pm_1 \rho_n] - [A^\pm_2, \rho_n A^\mp_1])\,,
\end{align}
where the upper (lower) sign has to be chosen if 
$\ket{s_\mathrm{in}} = \ket{\up}$ ($\ket{s_\mathrm{in}} = \ket{\down}$).
The second-order $z$-terms can be simplified as 
\begin{equation}
  \frac{\tau_1^2 g_1^2}{4} \diss(A^z_1)[\rho_n]
  + \frac{\tau_2^2 g_2^2}{4} \diss(A^z_2)[\rho_n]
  - \frac{\tau_1 g_1 \tau_2 g_2}{4} [A^z_2, [A^z_1, \rho_n]] 
  = \frac{1}{4} \diss(\tau_1 g_1 A^z_1 + \tau_2 g_2 A^z_2)[\rho_n]\,,
\end{equation}
while the $\pm$-terms can be rewritten as 
\begin{equation}
\begin{split}
  & \frac{\tau_1^2 g_1^2}{4} \diss(A^\pm_1)[\rho_n]
  + \frac{\tau_2^2 g_2^2}{4} \diss(A^\pm_2)[\rho_n]
  - \frac{\tau_1 g_1 \tau_2 g_2(1-2p)}{4} ([A^\mp_2, A^\pm_1 \rho_n] - 
  [A^\pm_2, \rho_n A^\mp_1]) 
  \\
  & \qquad = \frac{(1-2p)}{4} \diss(\tau_1 g_1 A^\pm_1 + \tau_2 g_2 A^\pm_2)[\rho_n]
  \mp \frac{\tau_1 g_1 \tau_2 g_2 (1-2p)}{8} ([A^-_2 A^+_1, \rho_n] + 
  [\rho_n, A^-_1 A^+_2])\\
  & \qquad\qquad
  + \frac{\tau_1^2 g_1^2 p}{2} \diss(A^\pm_1) + \frac{\tau_2^2 g_2^2 p}{2} \diss(A^\pm_2)\,.   
\end{split}
\end{equation}
Thus, after the first step of the protocol, the nuclei's reduced density matrix evolves as
\begin{equation}
  \begin{split}
    \rho_{n+1} & \simeq \rho_n - \frac{i}{2} \left[\mu_1 A^z_1 + \nu_2 A^z_2, \rho_n\right]
    + \frac{1}{4} \mathcal{D}\left(\mu_1 A^z_1 + \nu_2 A^z_2\right)[\rho_n] \\
    & \qquad + \frac{(1 - 2p)}{4} \mathcal{D}\left(\mu_1 A^+_1 + \nu_2 A^+_2\right)[\rho_n]
    - \frac{\mu_1\nu_2(1 - 2p)}{8}\left(\left[A^-_2A^+_1,\rho_n\right] + [\rho_n,A^-_1A^+_2]\right) \\
    & \qquad + \frac{\mu_1^2 p}{2} \mathcal{D}\left(A^+_1\right)[\rho_n]
    + \frac{\nu_2^2 p}{2} \mathcal{D}\left(A^+_2\right)[\rho_n] \,,
  \end{split} 
\end{equation}
while after the second step, the nuclei's reduced density matrix evolves as
\begin{equation}
  \begin{split}
    \rho_{n+2} & \simeq \rho_{n+1} + \frac{i}{2} \left[\nu_1 A^z_1 + \mu_2 A^z_2, \rho_{n+1}\right]
    + \frac{1}{4} \mathcal{D}\left(\nu_1 A^z_1 + \mu_2 A^z_2\right)[\rho_{n+1}] \\
    & \qquad + \frac{(1 - 2p)}{4} \mathcal{D}\left(\nu_1 A^-_1 + \mu_2 A^-_2\right)[\rho_{n+1}]
    + \frac{\nu_1\mu_2(1 - 2p)}{8}\left(\left[A^-_2A^+_1,\rho_{n+1}\right] + [\rho_{n+1},A^-_1A^+_2]\right) \\
    & \qquad + \frac{\nu_1^2 p}{2} \mathcal{D}\left(A^-_1\right)[\rho_{n+1}]
    + \frac{\mu_2^2 p}{2} \mathcal{D}\left(A^-_2\right)[\rho_{n+1}] \,.
  \end{split} 
\end{equation}
The concatenation of these two steps yields:
% \begin{equation}
%   \begin{split}
%     \rho_{n + 2} & \simeq \rho_n 
%     - \frac{i}{2} [\mu_1 A^z_1 + \nu_2 A^z_2, \rho_n]
%     + \frac{i}{2} [\nu_1 A^z_1 + \mu_2 A^z_2, \rho_n] 
%     \\ & \qquad
%     + \frac{1}{4} [\nu_1 A^z_1 + \mu_2 A^z_2, [\mu_1 A^z_1 + \nu_2 A^z_2, \rho_n]]
%     + \frac{1}{4} \diss (\mu_1 A^z_1 + \nu_2 A^z_2)[\rho_n]
%     + \frac{1}{4} \diss (\nu_1 A^z_1 + \mu_2 A^z_2)[\rho_n] 
%     \\ & \qquad
%     + \frac{(1-2p)}{4} \diss (\mu_1 A^+_1 + \nu_2 A^+_2)[\rho_n]
%     + \frac{(1-2p)}{4} \diss (\nu_1 A^-_1 + \mu_2 A^-_2)[\rho_n] \\
%     & \qquad + \frac{\nu_1^2 p}{2} \mathcal{D}\left(A^-_1\right)[\rho_n]
%     + \frac{\mu_2^2 p}{2} \mathcal{D}\left(A^-_2\right)[\rho_n]
%     + \frac{\mu_1^2 p}{2} \mathcal{D}\left(A^+_1\right)[\rho_n]
%     + \frac{\nu_2^2 p}{2} \mathcal{D}\left(A^+_2\right)[\rho_n] \\
%     & \qquad + \frac{(\mu_2\nu_1 - \mu_1\nu_2)(1 - 2p)}{8} 
%     \left(\left[A^-_2A^+_1,\rho_n\right] + [\rho_n,A^-_1A^+_2]\right)
%     \,.
%   \end{split} \label{eq:1stparttocho}
% \end{equation}
% Rearranging the terms in the second line of Eq.~\eqref{eq:1stparttocho}, we 
% finally obtain for the first part of the protocol:
\begin{equation}
  \begin{split}
    \rho_{n+2} & \simeq \rho_n 
    + \frac{i}{2}\left[(\nu_1 - \mu_1)A^z_1 + (\mu_2 - \nu_2)A^z_2, \rho_n\right] 
    + \frac{1}{4} \mathcal{D}\left((\nu_1-\mu_1)A^z_1 + (\mu_2-\nu_2)A^z_2\right)[\rho_n] \\ 
    & \qquad + \frac{(1 - 2p)}{4} \mathcal{D}\left(\nu_1A^-_1 + \mu_2A^-_2\right)[\rho_n]
    + \frac{(1 - 2p)}{4} \mathcal{D}\left(\mu_1 A^+_1 + \nu_2 A^+_2\right)[\rho_n] \\
    & \qquad + \frac{\nu_1^2 p}{2} \mathcal{D}\left(A^-_1\right)[\rho_n]
    + \frac{\mu_2^2 p}{2} \mathcal{D}\left(A^-_2\right)[\rho_n]
    + \frac{\mu_1^2 p}{2} \mathcal{D}\left(A^+_1\right)[\rho_n]
    + \frac{\nu_2^2 p}{2} \mathcal{D}\left(A^+_2\right)[\rho_n] \\
    & \qquad + \frac{(\mu_2\nu_1 - \mu_1\nu_2)(1 - 2p)}{8} 
    \left(\left[A^-_2A^+_1,\rho_n\right] + [\rho_n,A^-_1A^+_2]\right)
    \,.
  \end{split}
\end{equation}

For the second part, we first work out the concatenation of $m$ steps
with $\ket{s_\mathrm{in}} = \ket{\down}$, $\tau_1 g_1 = (\mu_1 - \nu_1)/k$ and 
$\tau_2 g_2 = 0$. Up to second order we find
\begin{equation}
  \rho_{n + 2 + m} \simeq \rho_{n + 2} 
  + \frac{i(\mu_1 - \nu_1)}{2k} c_m [A^z_1, \rho_{n + 2}]
  + \frac{(\mu_1 - \nu_1)^2}{4k^2} d_m \diss(A^z_1)[\rho_{n + 2}]
  + \frac{(\mu_1 - \nu_1)^2}{4k^2} c_m \diss(A^-_1)[\rho_{n + 2}] \,,
\end{equation}
where the coefficients obey a simple recurrence relation  
$c_m = c_{m - 1} + 1$ and $d_m = d_{m - 1} + 1 + 2c_{m-1}$ for $m > 1$, and 
$c_1 = d_1 = 1$. Therefore, the concatenation of $k$ such steps leads to
\begin{equation}
  \rho_{n + 2 + k} \simeq \rho_{n + 2} 
  + \frac{i(\mu_1 - \nu_1)}{2} [A^z_1, \rho_{n + 2}]
  + \frac{(\mu_1 - \nu_1)^2}{4} \diss(A^z_1)[\rho_{n + 2}]
  + \frac{(\mu_1 - \nu_1)^2}{4k} \diss(A^-_1)[\rho_{n + 2}] \,.
\end{equation}
Similarly, the concatenation of $k$ steps with $\ket{s_\mathrm{in}} = 
\ket{\up}$, $\tau_1 g_1 = 0$ and $\tau_2 g_2 = (\mu_2 - \nu_2)/k$, produces
\begin{equation}
  \rho_{n + 2 + 2k} \simeq \rho_{n + 2 + k} 
  - \frac{i(\mu_2 - \nu_2)}{2} [A^z_2, \rho_{n + 2 + k}]
  + \frac{(\mu_2 - \nu_2)^2}{4} \diss(A^z_2)[\rho_{n + 2 + k}]
  + \frac{(\mu_2 - \nu_2)^2}{4k} \diss(A^+_2)[\rho_{n + 2 + k}] \,.
\end{equation}
Thus, the concatenation of $k$ steps of the first kind, followed by $k$
steps of the second kind results in 
% \begin{equation}
%   \begin{split}
%     \rho_{n + 2 + 2k} & \simeq \rho_{n + 2} 
%     + \frac{i(\mu_1 - \nu_1)}{2} [A^z_1, \rho_{n + 2}] - \frac{i(\mu_2 - \nu_2)}{2} [A^z_2, \rho_{n + 2}]
%     \\ & \quad
%     + \frac{1}{4} [(\mu_2 - \nu_2)A^z_2, [(\mu_1 - \nu_1)A^z_1, \rho_{n + 2}]]
%     + \frac{(\mu_1 - \nu_1)^2}{4} \diss(A^z_1)[\rho_{n + 2}]
%     + \frac{(\mu_2 - \nu_2)^2}{4} \diss(A^z_2)[\rho_{n + 2}] 
%     \\ & \quad
%     + \frac{(\mu_1 - \nu_1)^2}{4k} \diss(A^-_1)[\rho_{n + 2}] 
%     + \frac{(\mu_2 - \nu_2)^2}{4k} \diss(A^+_2)[\rho_{n + 2}] \,,
%   \end{split}
% \end{equation}
% which can be simplified as
\begin{equation}
  \begin{split} 
    \rho_{n+2+2k} & \simeq \rho_{n+2} - \frac{i}{2}\left[(\nu_1 - \mu_1)A^z_1 + (\mu_2 - \nu_2)A^z_2, \rho_{n+2}\right]
    + \frac{1}{4} \mathcal{D}\left((\nu_1- \mu_1) A^z_1 + (\mu_2 - \nu_2) A^z_2\right)[\rho_{n+2}] \\
    & \qquad + \frac{(\mu_2 - \nu_2)^2}{4k} \mathcal{D}\left(A^+_2\right)[\rho_{n+2}]
    + \frac{(\mu_1 - \nu_1)^2}{4k} \mathcal{D}\left(A^-_1\right)[\rho_{n+2}]
    \,.
  \end{split} \label{eq:secondpart}
\end{equation}

Last, putting together the equations for the first and second part of the protocol, all $z$-terms cancel and we get 
\begin{equation}
  \begin{split}
    \Delta \rho_n & \simeq \frac{(1 - 2p)}{4} \mathcal{D}\left(\nu_1A^-_1 + \mu_2A^-_2\right)[\rho_n]
    + \frac{(1 - 2p)}{4} \mathcal{D}\left(\mu_1 A^+_1 + \nu_2 A^+_2\right)[\rho_n] \\
    & \qquad + \left[\frac{(\mu_1 - \nu_1)^2}{4k} + \frac{\nu_1^2 p}{2}\right] 
    \mathcal{D}\left(A^-_1\right)[\rho_n]
    + \left[\frac{(\mu_2 - \nu_2)^2}{4k} + \frac{\nu_2^2 p}{2}\right] \mathcal{D}\left(A^+_2\right)[\rho_n] \\
    & \qquad + \frac{\mu_2^2 p}{2} \mathcal{D}\left(A^-_2\right)[\rho_n]
    + \frac{\mu_1^2 p}{2} \mathcal{D}\left(A^+_1\right)[\rho_n] + \frac{(\mu_2\nu_1 - \mu_1\nu_2)(1 - 2p)}{8} 
    \left(\left[A^-_2A^+_1,\rho_n\right] + [\rho_n,A^-_1A^+_2]\right)
    \,,
  \end{split}\label{eq:megamastereq}
\end{equation}
where we have defined $\Delta\rho_n \equiv \rho_{n + 2 + 2k} - \rho_n$.
Eq.~\eqref{eq:mastereq} of the
main text corresponds to the case where $p=0$, $\mu_1=\mu_2=\mu$ and $\nu_1=\nu_2=\nu$.

\section{Fidelities for homogeneous systems}

\begin{figure}[!htb]
  \includegraphics{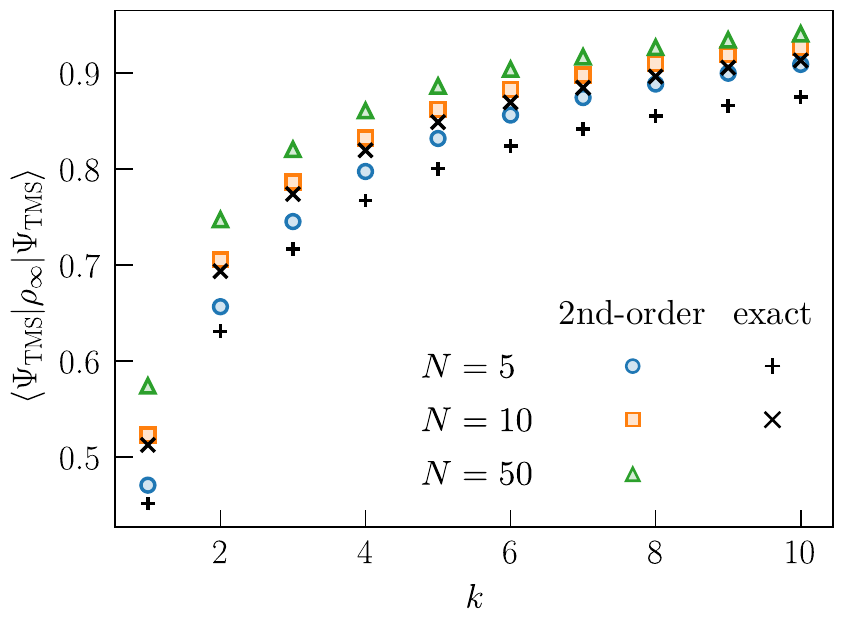}
  \caption{Fidelity of the steady state with respect to the two-mode spin squeezed state
  for an homogeneous system of spin-1/2 nuclei as a function of the number 
  of repetitions $k$ performed in the last two steps of the protocol. The 
  parameters of the system are $N_1=N_2=N$, $\mu=0.5/\sqrt{N}$ and 
  $\nu=0.8\mu$. The steady state has been computed using the approximate 
  2nd-order master equation, Eq.~\eqref{eq:mastereq} in the main text, and 
  also using the exact time evolution superoperator, Eq.~\eqref{eq:exactevol}
  in the main text. 
  \label{fig:fidelities}}
\end{figure}

In Fig.~\ref{fig:fidelities}, we show the fidelity of the steady 
state with respect to the two-mode spin squeezed state
[Eq.~\eqref{eq:psi_ideal} in the main text] as a function of the number of steps in the cycle. The system consists of two identical homogeneous ensembles of spin-1/2 nuclei. The initial state belongs to 
the subspace with maximal total angular momenta $J_j=N_j/2$. As expected, 
the fidelity increases for increasing $k$. 

\begin{figure}[!htb]
  \centering
  \includegraphics{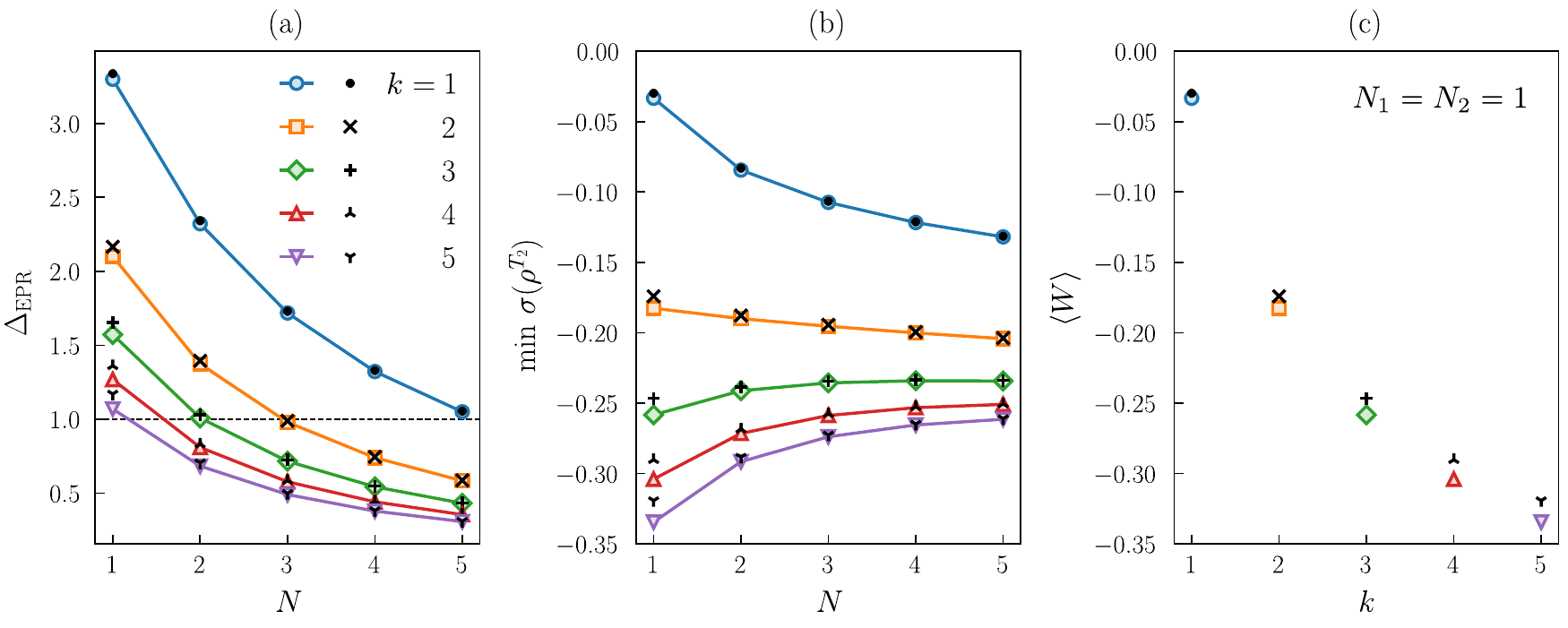}
  \caption{EPR uncertainty (a), minimum eigenvalue of the partial 
  transpose (b), and entanglement witness based on the singlet occupation (c), 
  for the steady state of a homogeneous system of spin-1/2
  nuclei with parameters: $N_1=N_2=N$, $\mu=0.1/\sqrt{N}$, and $\nu=0.8\mu$. 
  For small $k$ and $N$ the EPR criterion fails to detect entanglement, 
  although, the negativity of the partial transpose shows that the system is 
  indeed entangled. The steady state has been computed using the approximate 
  2nd-order master equation (filled symbols), Eq.~\eqref{eq:mastereq} in the 
  main text, and also using the exact time evolution superoperator (black 
  symbols), Eq.~\eqref{eq:exactevol} in the main text.}
  \label{fig:smallk}
\end{figure}

For small systems with $N_j<10$ the EPR uncertainty may not detect 
entanglement if $k$ is too small, see Fig.~\ref{fig:smallk}(a). Of course, 
making $k$ larger brings $\Delta_\mathrm{EPR}$ closer to the value in the 
ideal state, which does show entanglement irrespective of the number of 
nuclei considered. Nonetheless, according to the PPT criterion \cite{guhne2009}
the system is indeed entangled for all values of $k$ and $N$, see 
Fig.~\ref{fig:smallk}(b). But, how can this entanglement be detected 
experimentally? An entanglement witness for the simplest case of just two
qubits ($N_1=N_2=1$, $I=1/2$) can be constructed based on the following 
inequality, valid for separable states \cite{thekkadath2016},
\begin{equation}
  p_S\leq \frac{1-m^2}{2}\,.
\end{equation}
Here, $\bm{m}\equiv\mean{\bm{J}}$ ($m\equiv \norm{\bm m}$) is the expected 
value of the total spin, in this case $\bm{J}=\bm{I}_{11} + \bm{I}_{12}$, and 
$p_S\equiv\bra{\Psi_-}\rho\ket{\Psi_-}$ is the singlet occupation. Since 
$(1-m^2)\leq 1$, we have that $p_S>1/2$ implies that the two spins are 
entangled. Thus, we define the following entanglement witness
\begin{equation}
  W=\frac{1}{2}-\ket{\Psi_-}\bra{\Psi_-}=\frac{1}{2}(\bm{J}^2 - 1)
  =\frac{1}{4}+\bm{I}_{11}\cdot\bm{I}_{12}\,,
\end{equation}
such that $\mean{W}<0$ for entangled states. In Fig.~\ref{fig:smallk}(c) we 
show that indeed this entanglement witness is able to detect the entanglement 
produced by our protocol.

\section{Holstein-Primakoff transformation \label{sec:appendix_HolsteinPrimakoff}}

\begin{figure}[!htb]
  \centering
  \includegraphics{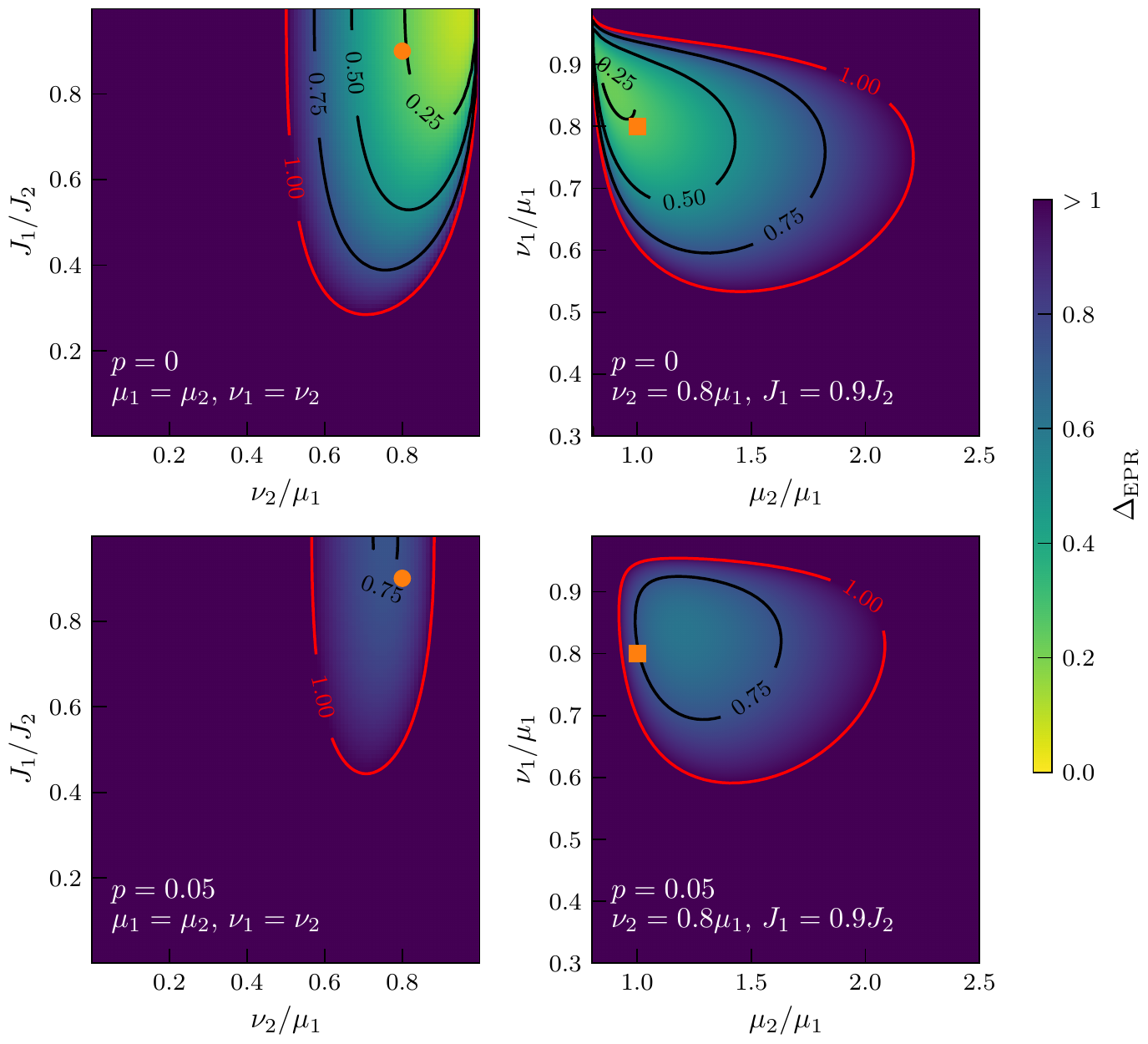}
  \caption{Dependence of the EPR uncertainty as given by the HP approximation on various parameters. In all the plots shown, $k=1$, $r_1=0.9$ and $r_2=0.8$. Other fixed parameters have the values indicated in each plot. The orange dots in the panels of the left column correspond to the fixed values of $N_1/N_2$ and $\nu_2/\mu_1$ used in the other two 
  panels. The orange squares in the panels of the right column correspond to the ``ideal'' choice $\mu_1=\mu_2$ and $\nu_1=\nu_2$.}
  \label{fig:EPRgeneral}
\end{figure}

In the limit of large total angular momenta and large polarizations, we can study the nuclear dynamics analytically replacing the collective total angular momentum operators by bosonic operators, $A^-_1\to\sqrt{r_1}a^\dag_1$ ($A^+_1\to\sqrt{r_1}a_1$) and $A^+_2\to\sqrt{r_2}a^\dag_2$ ($A^-_2\to\sqrt{r_2}a_2$) in Eq.~\eqref{eq:megamastereq}. Doing so, defining $\tilde \mu_j \equiv \mu_j \sqrt{r_j}$ and $\tilde \nu_j \equiv \nu_j \sqrt{r_j}$, we obtain the following master equation
\begin{equation}
  \begin{split}
    \Delta \rho_n & \simeq \frac{(1 - 2p)}{4} \mathcal{D}\big(\tilde\nu_1 a^\dag_1 + \tilde \mu_2 a_2\big)[\rho_n]
    + \frac{(1 - 2p)}{4} \mathcal{D}\big(\tilde \mu_1 a_1 + \tilde \nu_2 a^\dag_2\big)[\rho_n] \\
    & \qquad + \left[\frac{(\tilde\mu_1 - \tilde\nu_1)^2}{4k} + \frac{\tilde \nu_1^2 p}{2}\right] 
    \mathcal{D}\big(a^\dag_1\big)[\rho_n]
    + \left[\frac{(\tilde \mu_2 - \tilde \nu_2)^2}{4k} + \frac{\tilde \nu_2^2 p}{2}\right] \mathcal{D}\big(a^\dag_2\big)[\rho_n] \\
    & \qquad + \frac{\tilde\mu_2^2 p}{2} \mathcal{D}(a_2)[\rho_n]
    + \frac{\tilde \mu_1^2 p}{2} \mathcal{D}(a_1)[\rho_n] + \frac{(\tilde\mu_2\tilde\nu_1 - \tilde\mu_1\tilde\nu_2)(1 - 2p)}{8} 
    \big([a_2a_1,\rho_n] + [\rho_n,a^\dag_1a^\dag_2]\big)
    \,.
  \end{split}
\end{equation}
From this master equation, using the identities 
\begin{gather}
  \tr\left\{B[A,\rho]\right\} = \mean{[B,A]}\,,\\
  \tr\left\{B\diss(A)[\rho]\right\} = \frac{1}{2}
  \mean{[A^\dagger,B]A + A^\dagger[B,A]}\,, 
\end{gather}
one can obtain the equation of motion of any operator expressed in terms of $\{a_j,a^\dag_j\}_{j=1,2}$. Due to the bosonic commutation relations, the equation of motion of any product of these bosonic operators will only depend on products of the same or lower order. Hence, to find the steady-state expectation value of any set of monomials in these bosonic variables we will just have to solve a finite linear system of equations. We are interested in computing the EPR uncertainty, which depends on the absolute value polarizations $\abs{\mean{J^z_j}}\simeq J_j - \mean{a^\dag_j a_j}$ and variances
\begin{multline}
  \var(J^{x/y}_1 + J^{x/y}_2) \simeq \frac{J_1 + J_2}{2} 
  \pm \frac{J_1}{2} \mean{a^\dagger_1 a^\dagger_1 \pm 2 a^\dagger_1 a_1 + a_1 a_1}
  \pm \frac{J_2}{2} \mean{a^\dagger_2 a^\dagger_2 \pm 2 a^\dagger_2 a_2 + a_2 a_2} \\
  + \sqrt{J_1J_2} \mean{a^\dagger_1 a^\dagger_2 \pm a^\dagger_1 a_2 \pm a^\dagger_2 a_1 + a_1 a_2}
  \mp \left(\sqrt{\frac{J_1}{2}} \mean{a^\dagger_1 \pm a_1} \pm
  \sqrt{\frac{J_2}{2}} \mean{a^\dagger_2 \pm a_2} \right)^2 \,.
\end{multline}
Thus, we have to compute the steady-state expectation values of all monomials up to second order. Their equations of motion are:
\begin{gather}
  \Delta \mean{a_j} = \kappa_j \mean{a_j} + \delta_{j,2}\epsilon \mean{a^\dag_1} \,,\\
  \Delta \mean{a_j^2} = 2\kappa_j \mean{a_j^2} + \delta_{j,2} 2\epsilon \mean{a^\dagger_1 a_2} \,,\\
  \Delta \mean{a^\dagger_j a_j} = 2\kappa_j \mean{a^\dagger_j a_j} + 2\kappa_j + \frac{\tilde\mu_j^2}{4} 
  +\delta_{j,2} \epsilon \mean{a^\dagger_1 a^\dagger_2 + a_1 a_2} \,, \\
  \Delta \mean{a_1 a_2} = (\kappa_1 + \kappa_2) \mean{a_1 a_2} + \epsilon \mean{a^\dagger_1 a_1} 
  - \frac{(1 - 2p)\tilde\mu_2\tilde\nu_1}{4} \,,\\
  \Delta \mean{a^\dagger_1 a_2} = (\kappa_1 + \kappa_2) \mean{a^\dagger_1 a_2} 
  + \epsilon \mean{a^\dagger_1 a^\dagger_1} \,.
\end{gather}
where we have defined 
\begin{equation}
  \kappa_j = \frac{(\tilde\nu_j - \tilde\mu_j)}{8} \left[\tilde\nu_j + \tilde\mu_j + \frac{(\tilde\nu_j - \tilde\mu_j)}{k}\right]
  \,,\quad
  \epsilon = \frac{(1 - 2p)(\tilde\mu_1\tilde\nu_2 - \tilde\mu_2\tilde\nu_1)}{4} \,.
\end{equation}
The adjoint operators evolve according to $\Delta X^\dagger = \left(\Delta X\right)^\dagger$. By definition, the steady-state expectation value of an operator $X$, $\mean{X}_\infty$, satisfies $\Delta\mean{X}_\infty = 0$. For the operators we are considering, the only non-zero expectation values turn out to be
\begin{gather}
  \mean{a^\dagger_1 a_1}_\infty = -1 - \frac{\tilde\mu_1^2}{8\kappa_1} \,, \\
  \mean{a_1 a_2}_\infty = \mean{a^\dagger_1 a^\dagger_2}_\infty = 
  \frac{1}{\kappa_1 + \kappa_2}\left[\epsilon\left(1 + \frac{\tilde\mu_1^2}{8\kappa_1}\right)
  + \frac{(1 - 2p)\tilde\mu_2\tilde\nu_1}{4}\right] \,, \\
  \mean{a^\dagger_2 a_2}_\infty = - 1 - \frac{\tilde\mu_2^2}{8\kappa_2} 
  - \frac{\epsilon}{\kappa_2(\kappa_1 + \kappa_2)}
  \left[\epsilon\left(1 + \frac{\tilde\mu_1^2}{8\kappa_1}\right)
  + \frac{(1 - 2p)\tilde\mu_2\tilde\nu_1}{4}\right]
\end{gather}
Since many of the expectation values vanish, the formulas for the variances can be simplified as
\begin{equation}
  \var(J^x_1 + J^x_2)_\infty = \var(J^y_1 + J^y_2)_\infty = \frac{J_1 + J_2}{2} + J_1\mean{a^\dagger_1 a_1}
  + J_2\mean{a^\dagger_2 a_2} + 2\sqrt{J_1 J_2}\mean{a_1 a_2} \,.
\end{equation}
Substituting these back into the formulas for $\abs{\mean{J^z_j}}$ and $\var(J^{x/y}_1 + J^{x/y}_2)$, we can obtain an analytical expression for $\EPR$ in the steady state. 
In Eq.~\eqref{eq:HPDeltaEPR} of the main text we show the analytical formula obtained for $\EPR$ in the case of an ideal protocol ($p=0$, $\mu_1=\mu_2=\mu$ and $\nu_1=\nu_2=\nu$). In Fig.~\ref{fig:EPRgeneral} we show the values of $\EPR$ as a function of various parameters [cf. Fig.~\ref{fig:DeltaEPR_and_ss}(a) of the main text]. In the light of it we conclude that, at least for homogeneous ensembles, the protocol is able to produce entanglement for a wide range of values of the dwell times. 
% Furthermore, our original choice for the dwell times ($\mu_1=\mu_2$ and $\nu_1=\nu_2$) based on the intuition drawn from previous works (Refs.~\cite{muschik2011,schuetz2013,benito2016}) is not necessarily optimal. 
These results also demonstrate that the protocol is robust against certain amount of decoherence during electron transport. 

\begin{figure}[!htbp]
  \includegraphics{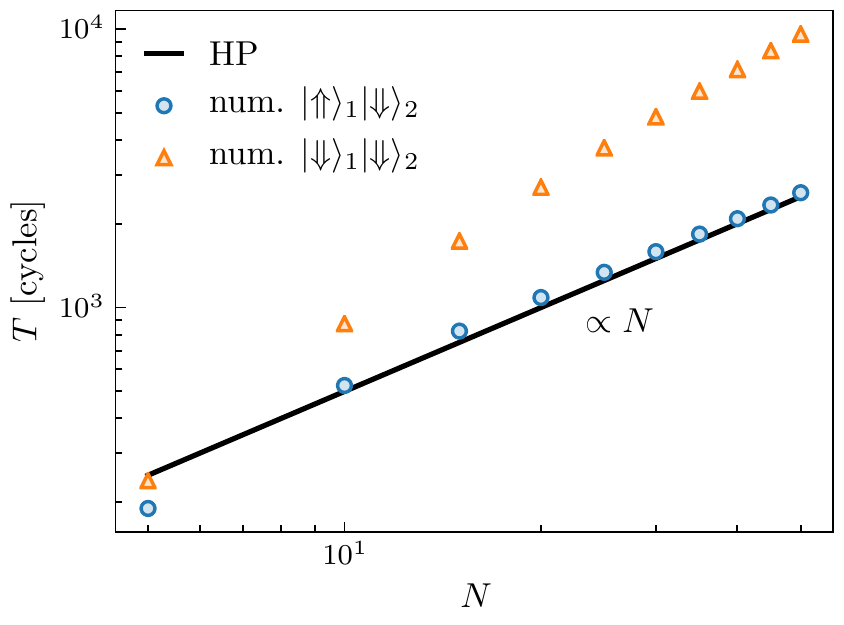}
  \caption{Stabilization time as a function of the number of nuclei 
  considered $N_1=N_2=N$ for an ideal protocol with parameters $\mu=0.5/\sqrt{N}$ and $\nu=0.8\mu$, performed in an homogeneous system of spin-1/2 nuclei. Filled symbols correspond to 
  the numerically obtained stabilization times for two different initial
  states, whereas the line corresponds to the formula $T=1/(2\kappa)$.
  \label{fig:lifetimes}}
\end{figure}

Moreover, from the corresponding EoMs, we can obtain the evolution of the expected number of bosons in each mode (and, therefore, the polarization of each ensemble). For an ideal protocol they are given by
\begin{equation}
  \mean{a^\dag_ja_j}_n = (1 - 2\kappa r_j)^n 
  \left(\mean{a^\dag_ja_j}_0 + \frac{\kappa - \eta}{2\kappa}\right)
  - \frac{\kappa - \eta}{2\kappa}\,,
\end{equation}
where we have defined $\kappa\equiv (\mu\nu - \nu^2)/4$ and $\eta\equiv \mu^2/4 - \kappa$. From these equations, we can deduce an analytical expression for the stabilization time
\begin{equation}
  T = -\frac{1}{\log(1-2\kappa r_j)} \simeq \frac{1}{2\kappa r_j}\,.
\end{equation}
In Fig.~\ref{fig:lifetimes}, we present results for the stabilization 
times of homogeneous systems. We show that this time depends on the 
initial state chosen, and for the fully polarized initial state 
$\ket{\Psi_\mathrm{FP}} = \ket{\Uparrow}_1\ket{\Downarrow}_2$ it follows 
the prediction of the HP approximation. 

\clearpage
\subsection{Mixed initial state}
\label{sec:mixedini}

In Fig.~\ref{fig:DeltaEPR_and_ss}(c) we consider a product initial state $\rho_0 = \rho_\mathrm{QD1}\otimes\rho_\mathrm{QD2}$, where the initial state of each ensemble is mixed $\rho_{\mathrm{QD}j}=\sum_J p_j(J)\rho_j(J)$. Here, $\rho_j(J)$ is a normalized density matrix that belongs to the $j$th-ensemble subspace of states with total angular momentum $J$. Normalization of $\rho_0$ requires $\sum_J p_j(J) = 1$. As explained in the main text, for homogeneous systems the steady state will be given by $\rho_\infty = \sum_{J_1,J_2} p_1(J_1)p_2(J_2) \rho_\infty(J_1, J_2)$, with $\rho_\infty(J_1,J_2)$ some steady-state within the $(J_1,J_2)$-subspace (the steady state is unique up to variations of the permutation quantum numbers). It can be computed numerically as the kernel of the Liouvillian corresponding to the evolution prescribed by Eq.~\eqref{eq:exactevol} or the master equation Eq.~\eqref{eq:mastereq}, or analytically using the HP transformation. However, this latter approach would only yield a good approximation if the initial distributions $p_j(J)$ are peaked towards higher angular momenta.

\begin{figure}[!htbp]
    \centering
    \includegraphics{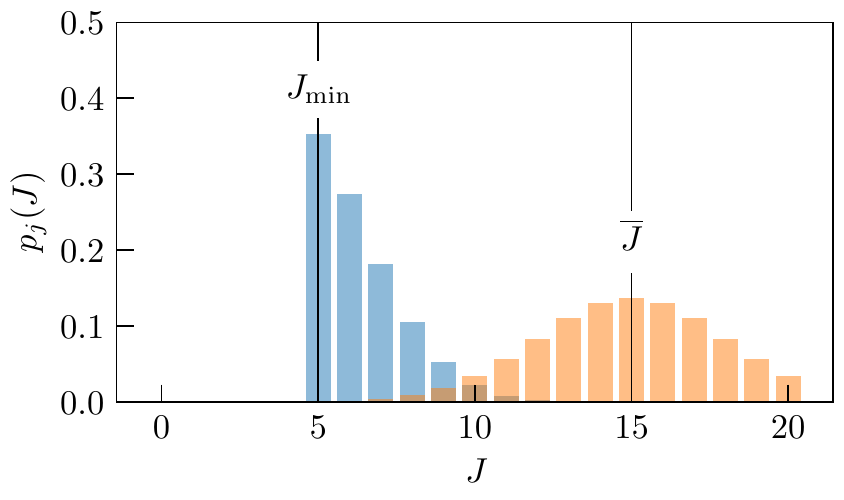}
    \caption{Equal-weight mixture distribution (blue bars) and Gaussian distribution with variance $w^2=3.0$ (orange bars) for an ensemble with $N=40$ spin-1/2 nuclei}
    \label{fig:distributions}
\end{figure}

In this manuscript we consider two different kinds of total angular momentum distributions. On the one hand, we consider an equal-weight mixture of all states with angular momenta $J_j\geq J^\mathrm{min}_j$. In this case the probabilities $p_j(J)$ are given, up to a normalization constant, by
\begin{equation}
    p_j(J) \propto \begin{cases} (2J + 1) D_j(J) \,, \text{ for } J \geq J^\mathrm{min}_j \\
    0 \,, \text{ for } J < J^\mathrm{min}_j\end{cases} \,,
\end{equation}
where 
\begin{equation}
  D_j(J) = \binom{N_j}{N_j/2 - J} - \binom{N_j}{N_j/2 - J - 1}
\end{equation}
corresponds to the degeneracy of the Dicke states for non-maximal total angular momentum $J$. Since the number of states with total angular momentum $J$ increases as $J$ decreases, this probability distribution is peaked at $J^\mathrm{min}_j$ and decays towards higher values of $J$. On the other hand, we have considered Gaussian distributions with mean value $\overline{J}$ and variance $w^2$, 
\begin{equation}
    p_j(J) \propto \exp\left(-\frac{\left(J - \overline{J}\right)^2}{2w^2}\right)\,.
\end{equation}
In Fig.~\ref{fig:distributions}, we compare both kinds of distributions.

\section{Non-linear equations of motion}

In this section we show how to obtain the EoMs, and asses the validity
of several different approximation schemes to close them. From the 
2nd-order master equation, Eq.~\eqref{eq:mastereq} in the main text, using the identity 
$\tr\left(B\diss(A)[\rho]\right) = \mean{[A^\dagger,B]A + A^\dagger[B,A]}/2$,
we obtain the following general EoMs, valid for any operator $X$, 
\begin{equation}
  \begin{split}
    \Delta\mean{X} & = \frac{\mu^2}{8} 
    \sum_{j=1,2} \mean{[[A^-_j, X], A^+_j] + [A^-_j, [X, A^+_j]]}
    + \frac{\mu\nu}{8} \sum_{j=1,2} 
    \mean{[[A^-_j, X], A^+_{3 - j}] + [A^-_j, [X, A^+_{3 - j}]]} 
    \\ & \quad
    + \frac{\nu^2 - \mu\nu}{4} \mean{[A^+_1, X]A^-_1 + A^+_1[X, A^-_1]
    + [A^-_2, X]A^+_2 + A^-_2[X, A^+_2]}\,.
  \end{split}
\end{equation}
Substituting $X$ by $\s{+}{ij}\s{-}{i'j'}$ and $\s{z}{ij}\s{z}{i'j'}$ we 
obtain the equations of motion presented in the main text, 
Eqs.~\eqref{eq:EoMgamma} and \eqref{eq:EoMlambda}. However, these 
equations are quite involved and do not allow a clear vision of the 
evolution of the system. For this, we will rather investigate first the 
polarization dynamics in each ensemble separately, and then investigate
the dynamics of the transverse spin variance.

\subsection{Polarization dynamics}
From the 2nd-order master equation, 
% using the identity $\tr\left(B\diss(A)\rho\right) = 
% \mean{[A^\dagger,B]A + A^\dagger[B,A]}/2$, 
we find the following general equation of motion
\begin{equation}
  \Delta\mean{X_1} = \frac{\eta - \kappa}{2}
  \mean{[A^+_1, X_1] A^-_1 + A^+_1 [X_1, A^-_1]}
  + \frac{\eta + \kappa}{2}
  \mean{[A^-_1, X_1] A^+_1 + A^-_1 [X_1, A^+_1]}\,,
  \label{eq:EoMX1}
\end{equation}
valid for any operator 
$X_1$ that acts non-trivially only on the nuclei belonging to the first 
ensemble. For an operator $X_2$ that acts non-trivially only on the nuclei 
belonging to the second ensemble, we obtain the same equation, replacing 
$A^\pm_1\to A^\mp_2$, or, equivalently, changing $\kappa\to -\kappa$ and the
subindex $1\to 2$. 

We can realize that if the coupling constants $\{g_{i2}\}$ and $\{g_{i1}\}$ 
are the same, that is, if the two ensembles are identical, the steady-state 
polarization of the second ensemble must be the same as that of the first, 
but with opposite sign. This can be seen applying a transformation 
$U=\prod_i\s{x}{i2}$, which transforms $A^\pm_2 \to \tilde{A}^{\mp}_2$, 
$J^z_2 \to -\tilde{J}^z_2$. Thus, the EoM for $\mean{\tilde{J}^z_2}$ would 
be the same as that for $\mean{J^z_1}$ (also with the same initial value if 
the initial state is $\ket{\Uparrow}_1\ket{\Downarrow}_2$). Undoing the 
transformation, the steady-state polarization in QD2 is 
$\mean{J^z_2}_\infty = -\mean{\tilde{J}^z_2}_\infty = -\mean{J^z_1}_\infty$.

To compute the polarization of the first ensemble, we particularize 
Eq.~\eqref{eq:EoMX1} for $X_1 = \s{z}{i1}$ ($i=1,\dots,N_1$), obtaining
\begin{equation}
  \Delta \mean{\s{z}{i}} = (\kappa - \eta) \sum_k g_k g_i 
  \left(\mean{\s{+}{i}\s{-}{k}} + \mean{\s{+}{k}\s{-}{i}}\right)
  + (\kappa + \eta) \sum_k g_k g_i
  \left(\mean{\s{-}{i}\s{+}{k}} + \mean{\s{-}{k}\s{+}{i}}\right) .
  \label{eq:pol1}
\end{equation}
Here, and in the following equations, we have dropped the QD index to avoid
cluttering. It should be understood that all operators and coupling 
constants refer to the first ensemble. As it turns out, the individual 
polarizations depend on all the matrix elements of the covariance matrix 
$\gamma_{ij} \equiv \mean{\s{+}{i}\s{-}{j}}$, which, in turn, depend on 
higher order correlations as 
\begin{align}
  \begin{split}
    \Delta\gamma_{ij} & = \frac{\eta - \kappa}{2} \sum_k g_k 
    \left(g_j\mean{\s{+}{i}[\s{+}{j},\s{-}{j}]\s{-}{k}} 
    + g_i\mean{\s{+}{k}[\s{+}{i},\s{-}{i}]\s{-}{j}}\right) 
    \\ & \quad
    + \frac{\eta + \kappa}{2} \sum_k g_k 
    \left(g_i\mean{[\s{-}{i},\s{+}{i}]\s{-}{j}\s{+}{k}} 
    + g_j\mean{\s{-}{k}\s{+}{i}[\s{-}{j},\s{+}{j}]}\right) ,
    \label{eq:movida}
  \end{split}
  \\
    & = \frac{\eta - \kappa}{2} \sum_k g_k 
    \left(g_j\mean{\s{+}{i}\s{z}{j}\s{-}{k}} 
    + g_i\mean{\s{+}{k}\s{z}{i}\s{-}{j}}\right) 
    - \frac{\eta + \kappa}{2} \sum_k g_k 
    \left(g_i\mean{\s{z}{i}\s{-}{j}\s{+}{k}} 
    + g_j\mean{\s{-}{k}\s{+}{i}\s{z}{j}}\right) .
\end{align}
Rewriting the terms in the second sum as those in the first, we obtain
\begin{equation}
  \Delta\gamma_{ij} = - \kappa \sum_k g_k \left( g_j 
  \mean{\s{+}{i}\s{z}{j}\s{-}{k}} 
  + g_i \mean{\s{+}{k}\s{z}{i}\s{-}{j}} \right) - (\kappa + \eta)
  \left[(g_i^2 + g_j^2)\gamma_{ij} - g_ig_j\lambda_{ij}\right] .
\end{equation}
where $\lambda_{ij} = \mean{\s{z}{i}\s{z}{j}}$.
Note that higher-order correlations can be simplified in case that some 
indices repeat,
\begin{equation}
  \s{+}{i}\s{z}{j}\s{-}{k}=\begin{cases}
    -\s{+}{i}\s{-}{k}\,, & \text{if $i=j$, or $j=k$;}\\
    (\s{z}{j}+\s{z}{i}\s{z}{j})/2\,, & \text{if $i=k\neq j$.}
  \end{cases}
\end{equation}
Therefore,
\begin{align}
  \Delta\gamma_{ii} & = \kappa \sum_{k\neq i} g_k g_i (\gamma_{ik} + \gamma_{ki})
  - 2\eta g_i^2 \gamma_{ii} + (\kappa + \eta) g_i^2 \,, 
  \label{eq:gammaii}\\
  \Delta\gamma_{ij} & = - \kappa \sum_{k\neq i, j} g_k \left(
  g_j \mean{\s{+}{i}\s{z}{j}\s{-}{k}} + g_i \mean{\s{+}{k}\s{z}{i}\s{-}{j}}
  \right) - \eta (g_i^2 + g_j^2)\gamma_{ij} + \eta g_i g_j \lambda_{ij} + 
- \kappa g_i g_j (\gamma_{ii} + \gamma_{jj} - 1)\,, \ \text{for $i\neq j$.}
  \label{eq:gammaij}
\end{align}
The EoMs for the $zz$-correlations can be obtained in an analogous way, 
\begin{equation}
  \begin{split}
    \Delta\lambda_{ij} & = 2\kappa \sum_{k\neq i,j} g_k 
    \left[g_j \left(\mean{\s{+}{j}\s{z}{i}\s{-}{k}}
  + \mean{\s{+}{k}\s{z}{i}\s{-}{j}}\right) + g_i \left(
  \mean{\s{+}{i}\s{z}{j}\s{-}{k}} + \mean{\s{+}{k}\s{z}{j}\s{-}{i}}\right)
  \right] \\
    & \quad + 4\eta g_i g_j (\gamma_{ij} + \gamma_{ji}) 
    - 2\eta (g_i^2 + g_j^2) \lambda_{ij} 
    + 2\kappa\left[g_i^2(2\gamma_{jj} - 1) + g_j^2(2\gamma_{ii} -1)\right] 
    , \ \text{for $i\neq j$.}
  \end{split} \label{eq:lambdaij}
\end{equation}
For the nuclei in the second ensemble we obtain equations similar to 
those presented so far, the only difference being the change of sign
$\kappa\to -\kappa$.

The different approximation schemes that we have considered 
are \cite{christ2007}:
\begin{enumerate}
  \item \emph{Spin-temperature:} Neglecting all correlations 
    $\mean{\s{+}{i}\s{-}{j}}\approx 0$ for $i\neq j$, Eq.~\eqref{eq:pol1}
    leads to an independent EoM for each nucleus
    \begin{equation}
      \Delta\mean{\s{z}{i}} = -2\eta g_i^2 \mean{\s{z}{i}}
      + 2\kappa g_i^2 \,.
      % \quad \Rightarrow \quad 
      % \mean{\s{z}{i}}_\infty = \frac{\kappa}{\eta}\,.
    \end{equation}
  \item \emph{Boson:} Deviations from the fully polarized state can be 
    approximated as bosonic excitations. Substituting spin operators by 
    bosonic operators,
    $\s{-}{i}\to a^\dagger_i$ and $\s{z}{i}\to 1 - 2 a^\dagger_i a_i$, for 
    the nuclei in the first ensemble, we can approximate 
    $[\s{+}{i},\s{-}{j}]\approx\delta_{ij}$ in Eq.~\eqref{eq:movida} 
    obtaining a closed set of equations:
    \begin{equation}
      \Delta\gamma_{ij} = - \kappa \sum_k g_k (g_j \gamma_{ik} 
      + g_i \gamma_{kj}) + (\eta + \kappa)g_ig_j\,.
    \end{equation}
    Note that within this approximation the polarization of each 
    nucleus in the first ensemble should be computed as 
    $\mean{\s{z}{i}} = 3 - 2\gamma_{ii}$. For the second ensemble 
    we substitute $\s{-}{i}\to a_i$ and $\s{z}{i} \to 2 a^\dagger_i a_i - 1$
    ($[\s{+}{i},\s{-}{j}]\approx -\delta_{ij}$), obtaining
    \begin{equation}
      \Delta\gamma_{ij} = - \kappa \sum_k g_k 
      (g_j \gamma_{ik} + g_i \gamma_{kj}) + (\eta - \kappa) g_i g_j \,.
    \end{equation}
    The polarization of the nuclei in the second ensemble can be computed 
    using the usual formula $\mean{\s{z}{i}} = 2\gamma_{ii} - 1$.

  % \item \emph{Boson:} Deviations from the fully polarized state can be 
  %   approximated as bosonic excitations. Substituting the spin operators by 
  %   bosonic operators $\s{-}{i}\to a^\dagger_i$, 
  %   $\s{z}{i}\to 1 - 2a^\dagger_ia_i$ for the nuclei in QD1; and 
  %   $\s{+}{i}\to a^\dagger_i$, $\s{z}{i}\to 2a^\dagger_ia_i - 1$ for the 
  %   nuclei in QD2, we obtain from Eq.~\eqref{eq:EoMX1} the following EoM
  %   \begin{equation}
  %     \Delta\mean{a^\dagger_i a_j} = \kappa \sum_k g_k 
  %     \big(g_j \mean{a^\dagger_i a_k} + g_i \mean{a^\dagger_k a_j}\big)
  %     + (\eta + \kappa) g_i g_j \,.
  %   \end{equation}
  %   The EoM for the nuclei in the second ensemble is exactly the same, the
  %   change in the sign of $\kappa$ gets compensated by the different 
  %   transformation of the ladder operators.
  %   The particular way in which we substitute spin operators by bosonic 
  %   operators is important. For example, if we substitute instead 
  %   $\s{+}{i}\to a^\dagger_i$, $\s{z}{i}\to 2a^\dagger_i a_i - 1$ for the
  %   nuclei in QD1, the value of $\mean{a^\dagger_i a_i}$ would be larger, 
  %   but this approximation scheme works best the smaller these expected 
  %   values are, so our initial choice is preferable for the initial state
  %   we are considering.
\end{enumerate}
For the rest of schemes, we use Eqs.~\eqref{eq:gammaii} and 
\eqref{eq:gammaij}, approximating
\begin{enumerate}[resume]
  \item \emph{Hartree:} $\mean{\s{+}{i}\s{z}{j}\s{-}{k}} \approx 
    (2\gamma_{jj} - 1)\gamma_{ik}$.
  \item \emph{Wick:} $\mean{\s{+}{i}\s{z}{j}\s{-}{k}} \approx 
    (2\gamma_{jj} - 1)\gamma_{ik} + 2\gamma_{ij}\gamma_{jk}$.
  \item \emph{Spin:} $\mean{\s{+}{i}\s{z}{j}\s{-}{k}} \approx 
    (2\gamma_{jj} - 1)\gamma_{ik} - 2\gamma_{ij}\gamma_{jk}$.
\end{enumerate}
We may approximate as well the $zz$-correlations 
$\mean{\s{z}{i}\s{z}{j}}=2\mean{\s{+}{i}\s{z}{j}\s{-}{i}} 
- \mean{\s{z}{j}}$, or use its own equation of motion, 
Eq.~\eqref{eq:lambdaij}.

\begin{figure}[!htb]
  \includegraphics{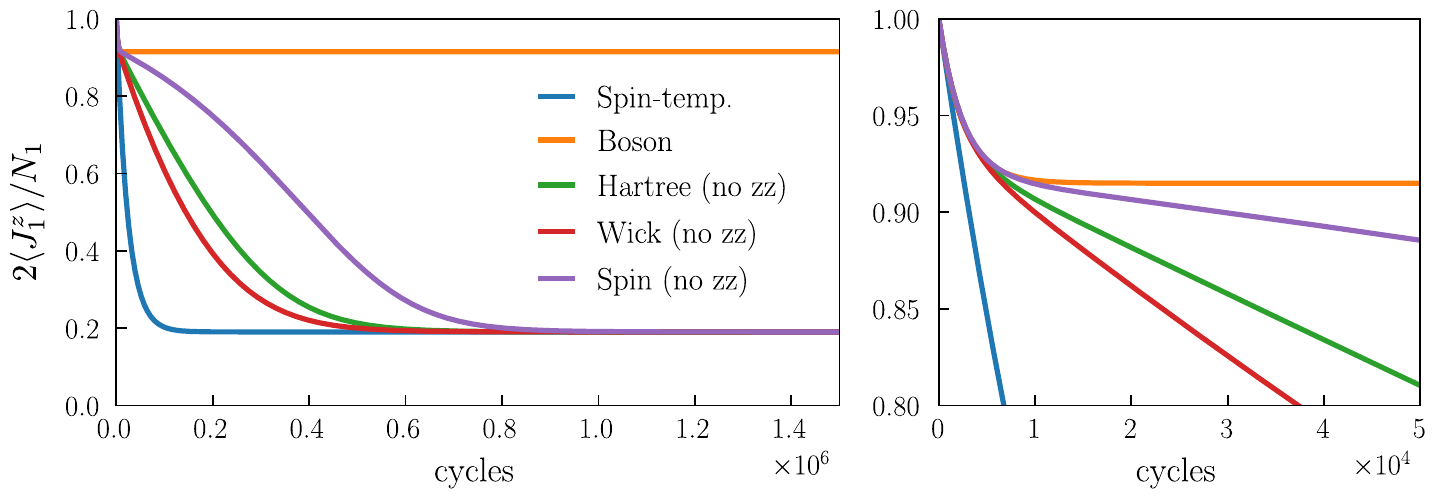}
  \caption{Polarization dynamics for the first ensemble as given by the 
  different approximation schemes. The plot on the right is just a zoom in 
  the short-term dynamics. The results correspond to an homogeneous 
  system with parameters: $N_1=50$, $\mu=0.5/\sqrt{N_1}$, and $\nu=0.8\mu$.
  The Hartree, Wick, and Spin approximations have been used without 
  including the EoMs for the $zz$-correlations. 
  \label{fig:polz_EoMhom_nozz}}
\end{figure}

\begin{figure}[!htb]
  \includegraphics{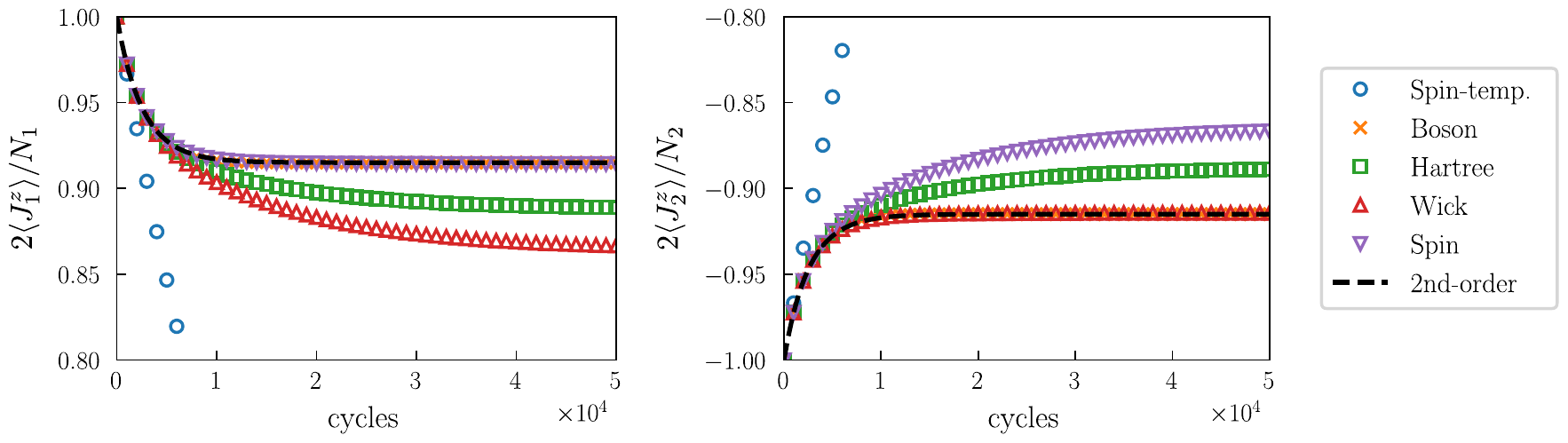}
  \caption{Polarization in the first (left) and second (right) ensembles 
  for an homogeneous system with $N_1=N_2=50$, $\mu=0.5/\sqrt{N_1}$, and 
  $\nu=0.8\mu$. The markers correspond to the different approximation 
  schemes 1--5, while the dashed line corresponds to the result given by
  the 2nd-order master equation, Eq.~\eqref{eq:mastereq} in the main 
  text. \label{fig:polz_EoMhom}}
\end{figure}

In Fig.~\ref{fig:polz_EoMhom_nozz} we show the polarization dynamics of the 
first ensemble, obtained using the different approximation schemes 1--5. 
The Hartree, Wick, and Spin approximations are implemented without 
including the EoMs for the $zz$-correlations. As can be appreciated, these
three approaches result in different dynamics that eventually converge
to the same steady-state value as that obtained in the Spin-temperature 
description. If we include the EoMs for the $zz$-correlations, then the 
results are much better, see Fig.~\ref{fig:polz_EoMhom}. Curiously enough, 
the best approximation for the first ensemble is the Spin approximation, 
while the best for the second is the Wick approximation. This can be 
understood as follows: In a rotated frame, 
we already established that the equation of motion of 
$\mean{\tilde{\sigma}^z_{i2}}$ is the same as that of $\mean{\s{z}{i1}}$.
Applying the Spin approximation in this rotated frame amounts to 
\begin{equation}
  \mean{\tilde{\sigma}^+_i\tilde{\sigma}^z_j\tilde{\sigma}^-_k} \approx
  \mean{\tilde{\sigma}^z_j} \mean{\tilde{\sigma}^+_i\tilde{\sigma}^-_k} 
  - 2 \mean{\tilde{\sigma}^+_i\tilde{\sigma}^-_j} 
  \mean{\tilde{\sigma}^+_j\tilde{\sigma}^-_k} \,,
  \quad \text{for $i\neq j$, $j\neq k$, and $i\neq k$.}
\end{equation}
Rotating back to the original frame we have
\begin{equation}
  - \mean{\s{-}{i}\s{z}{j}\s{+}{k}} \approx - \mean{\s{z}{j}}
  \mean{\s{-}{i}\s{+}{k}} - 2 \mean{\s{-}{i}\s{+}{j}}\mean{\s{-}{j}\s{+}{k}}
  \,, \quad \text{for $i\neq j$, $j\neq k$, and $i\neq k$.}
\end{equation}
Rearranging the operators, we then have
\begin{equation}
  \mean{\s{+}{k}\s{z}{j}\s{-}{i}} \approx \mean{\s{z}{j}}
  \mean{\s{+}{k}\s{-}{i}} + 2 \mean{\s{+}{j}\s{-}{i}}\mean{\s{+}{k}\s{-}{j}}
  \,, \quad \text{for $i\neq j$, $j\neq k$, and $i\neq k$,}
\end{equation}
which is nothing but the Wick approximation.
We also note that the Boson approximation seems to be fine. However, as we
show next, it is not so good for the variances.

\subsection{Variance dynamics}

Due to the symmetries of the protocol (we only send electrons polarized in 
the z-direction), we expect $\var(J^x_1 + J^x_2) = \var(J^y_1 + J^y_2)$. 
The variance of the total nuclear spin along the $x$-direction is
\begin{equation}
  \var(J^x_1 + J^x_2) = \frac{1}{4} \sum_{i,\,j,\,i',\,j'}
  \mean{\s{+}{ij}\s{+}{i'j'} + \s{+}{ij}\s{-}{i'j'} + \s{-}{ij}\s{+}{i'j'} 
  + \s{-}{ij}\s{-}{i'j'}} 
  - \frac{1}{4}\Bigg(\sum_{i,\,j} \mean{\s{+}{ij} + \s{-}{ij}}\Bigg)^2 .
\end{equation}
Note that $\mean{\s{-}{ij}} = \mean{\s{+}{ij}}^*$, 
$\mean{\s{-}{ij}\s{-}{i'j'}} = \mean{\s{+}{i'j'}\s{+}{ij}}^*$, and
$\mean{\s{+}{ij}\s{-}{i'j'}} = \mean{\s{+}{i'j'}\s{-}{ij}}^*$ ($z^*$ denotes
the complex conjugate of $z$). However, since the Liouvillian superoperator
and the initial nuclei's reduced density matrix are both real, the 
covariance matrix is symmetric at all times, $\mean{\s{+}{ij}\s{-}{i'j'}} = 
\mean{\s{+}{i'j'}\s{-}{ij}}$. This property of the covariance matrix is 
also maintained by the EoMs. Also, it can be shown that if 
$\mean{\s{+}{ij}}$ and $\mean{\s{+}{ij}\s{+}{i'j'}}$ are zero initially, 
they remain equal to zero throughout time evolution. With this assumption, 
noting that $\{\s{+}{ij}, \s{-}{ij}\}=1$, we have
% \begin{equation}
%   \var(J^x_1 + J^x_2) = \frac{1}{2}\sum_{\substack{i,\,j,\,i',\,j' \\ 
%   i,\,j \neq i',\,j'}} \gamma^{jj'}_{ii'} + \frac{N_1 + N_2}{4} \,.
% \end{equation}
\begin{equation}
  \var(J^x_1 + J^x_2) = \frac{1}{2}\sum_{i,\,j \neq i',\,j'} 
  \gamma^{jj'}_{ii'} + \frac{N_1 + N_2}{4} \,.
\end{equation}
We have worked out already the EoMs of $\gamma^{11}_{ij}$ and 
$\gamma^{22}_{ij}$ in the previous section, and found the best 
approximations to compute them, namely 
\begin{equation}
  \mean{\s{+}{i1}\s{z}{j1}\s{-}{k1}}\approx 
  \left(2\gamma^{11}_{jj} - 1\right)\gamma^{11}_{ik} - 2\gamma^{11}_{ij}
  \gamma^{11}_{jk}\,, \
  \mean{\s{+}{i2}\s{z}{j2}\s{-}{k2}}\approx 
  \left(2\gamma^{22}_{jj} - 2\right)\gamma^{22}_{ik} + 2\gamma^{22}_{ij}
  \gamma^{22}_{jk}\,, 
\end{equation}
for $i\neq j$, $j\neq k$, and $i\neq k$. However, the remaining covariance
matrix elements depend on higher-order correlations of spins belonging to
different ensembles, as can be seen in their EoMs
% . The EoMs of $\gamma^{12}_{ij}$ and 
% $\lambda^{12}_{ij}$ can be shown to be
\begin{align}
  \Delta\gamma^{12}_{ij} & = - \kappa \sum_{k\neq i} g_{k1}g_{i1}
  \mean{\s{+}{k1}\s{z}{i1}\s{-}{j2}} - \sum_{k\neq j} g_{k2}g_{j2}
  \mean{\s{+}{i1}\s{z}{j2}\s{-}{k2}} 
  - \eta \left(g_{i1}^2 + g_{j2}^2\right) \gamma^{12}_{ij}
  + \frac{\mu\nu}{4} g_{i1}g_{j2} \lambda^{12}_{ij} \,, \label{eq:gamma12}
  \\
  \begin{split}
  \Delta\lambda^{12}_{ij} & = 2\kappa \sum_{k\neq i} g_{k1}g_{i1} \left(
  \mean{\s{+}{i1}\s{z}{j2}\s{-}{k1}} + \mean{\s{+}{k1}\s{z}{j2}\s{-}{i1}}
  \right) - 2\kappa \sum_{k\neq j} g_{k2}g_{j2} \left(
  \mean{\s{+}{j2}\s{z}{i1}\s{-}{k2}} + \mean{\s{+}{k2}\s{z}{i1}\s{-}{j2}}
  \right) 
  \\
  & \quad -2\eta \left(g^2_{i1} + g^2_{j2}\right) \lambda^{12}_{ij}
  +\mu\nu g_{i1}g_{j2} \left(\gamma^{21}_{ji} + \gamma^{12}_{ij}\right)
  - 2\kappa\left[g^2_{j2}(2\gamma^{11}_{ii} - 1) 
    - g^2_{i1}(2\gamma^{22}_{jj} - 1)\right]\,.
  \end{split} \label{eq:lambda12}
\end{align}
Below, we discuss several approaches to close the EoMs and compute 
$\gamma^{12}_{ij}$:
\begin{enumerate}
  \item \emph{Spin-temperature:} In this approach we neglect all 
    correlations between different nuclei, so $\gamma^{jj'}_{ii'}\approx 0$,
    and the variance is simply given by
    $\var(J^x_1 + J^x_2)\approx (N_1 + N_2)/4$.
  \item \emph{Boson:} Approximating $[\s{+}{i1},\s{-}{j1}] \approx
    \delta_{ij}$, and $[\s{+}{i2},\s{-}{j2}] \approx -\delta_{ij}$, we 
    obtain directly from the 2nd-order master equation the following closed
    system of equations,
    \begin{equation}
      \Delta\gamma^{12}_{ij} = - \kappa 
      \left(\sum_k g_{k1}g_{i1}\gamma^{12}_{kj} 
      + \sum_k g_{k2}g_{j2}\gamma^{12}_{ik}\right) - \frac{\mu\nu}{4}
      g_{i1}g_{j2}\,.
    \end{equation}
  \item \emph{Hartree:} As can be appreciated, in 
    Fig.~\ref{fig:polz_EoMhom}, the Hartree approximation lies in between 
    the Spin and the Wick approximations. Thus, we may approximate 
    higher-order correlations that involve spins in different ensembles
    using the Hartree approximation.
  \item \emph{Half-Hartree:} Approximate 
    $\mean{\s{+}{ij}\s{z}{i'j'}\s{-}{kl}}$ using the Hartree approximation
    if $j\neq l$, and the Spin (Wick) approximation if $j=l=1$ ($j=l=2$).
  \item \emph{Majority:} Approximate higher-order correlations that involve spins in different ensembles using the Spin (Wick) approximation if the majority of operators belong to the first (second) ensemble.
  \item \emph{Z-rules:} Approximate $\mean{\s{+}{ij}\s{z}{i'j'}\s{-}{kl}}$
    using the Spin (Wick) approximation if $j'=1$ ($j'=2$).
\end{enumerate}

\begin{figure}[!htb]
  \includegraphics[width=\linewidth]{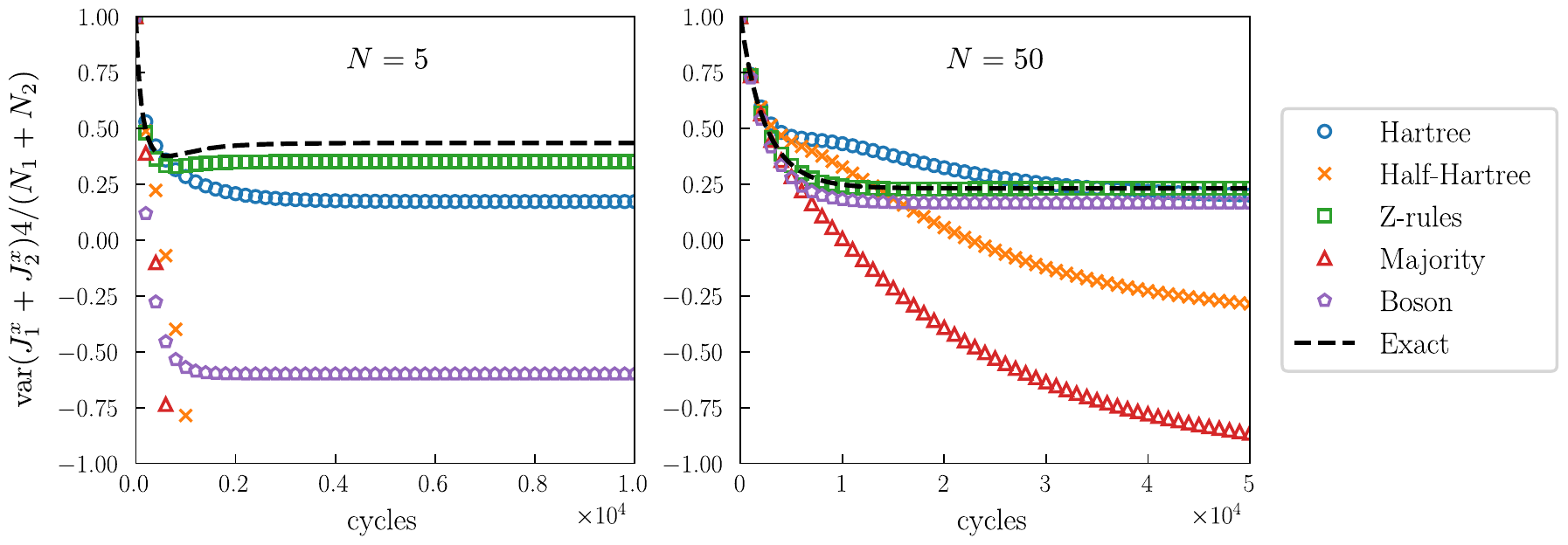}
  \caption{Variance of the total spin along a transverse direction for 
  homogeneous systems with $N_1=N_2=N$, $\mu=0.5/\sqrt{N}$, and 
  $\nu=0.8/\mu$. The markers correspond to the approximation schemes 2--6,
  while the dashed line corresponds to the result given by the 2nd-order
  master equation. \label{fig:variance_hom}}
\end{figure}

In Fig.~\ref{fig:variance_hom} we show the resulting dynamics of the 
transverse spin variance for an homogeneous system with ensemble sizes
$N_1=N_2=5$ (left), and $N_1=N_2=50$ (right). All approximations seem to 
work better for the larger system. Among all of them, the closest to the 
actual dynamics is the Z-rules approximation. 

\begin{figure}
  \includegraphics{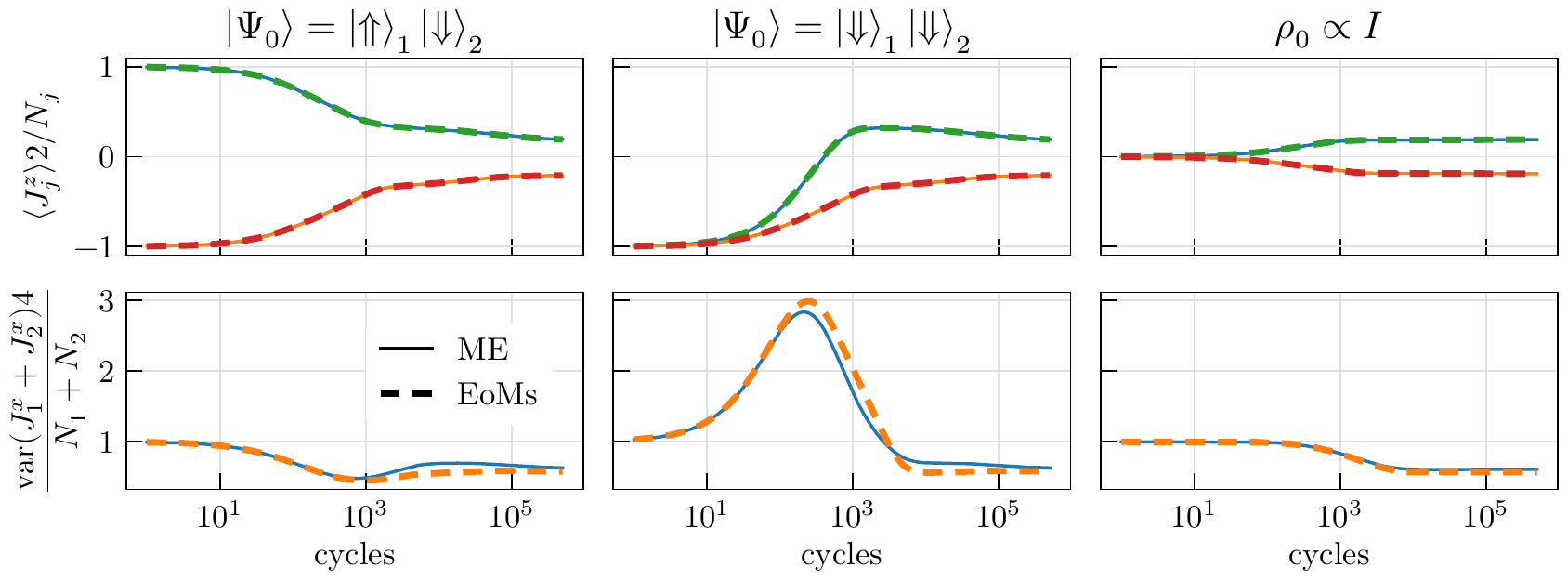}
  \caption{Evolution of the polarizations of each ensemble and variance of the total spin in the transverse plane
  for an inhomogeneous system with $N_1=N_2=5$. Each column corresponds to a different initial state ($\rho_0=\ket{\Psi_0}\bra{\Psi_0}$).
  Continuous lines show the results obtained using the 2nd-order master equation
  [Eq.~\eqref{eq:mastereq} in the main text], while dashed lines show the results given by the EoMs 
  [Eqs.~\eqref{eq:EoMgamma} and \eqref{eq:EoMlambda} in the main text]. The coupling constants are all different and they have a relative standard deviation   of 32\% for the first ensemble, and 51\% for the second; they are the same for all different initial states considered. The remaining parameters have values $\mu=0.5/\sqrt{N_1}$ and $\nu=0.8\mu$. 
  \label{fig:N5_inhom}}
\end{figure}

As a final check, we compare the dynamics given by the Z-rules approximation with the actual dynamics given by the second-order master equation for a small inhomogeneous system, see Fig.~\ref{fig:N5_inhom}. While the polarizations seem to be very well approximated by the EoMs, the variances are a bit underestimated. However, as we show in Fig.~\ref{fig:variance_hom}, calculations for homogeneous systems of different sizes suggest that this difference could be less significant the larger the size of the ensembles. Remarkably, all different initial states considered converge to the same steady state.

% \begin{figure}
% %   \includegraphics{5dots_inhom_log_newWitness.pdf}
%   \includegraphics{2ndorder_vs_EOM_mixed.pdf}
%   \caption{Evolution of the generalized polarization of the first ensemble
%   and transverse spin variance, divided by their values in the fully 
%   polarized state, 
%   $\mean{A^{(z,2)_1}}_\mathrm{FP}=\var(A^x_1+A^x_2)_\mathrm{FP}=1/2$,
%   for an inhomogeneous system with $N_1=N_2=5$. 
%   The initial state is the fully polarized state
%   $\rho_0 = \ket{\Psi_\mathrm{FP}} \bra{\Psi_\mathrm{FP}}$. 
%   % The magnitudes
%   % plotted are divided by their values in the initial state
%   % $\mean{A^{(z,2)_1}}_\mathrm{FP}=\var(A^x_1+A^x_2)_\mathrm{FP}=1/2$.
%   Lines show the results obtained using the 2nd-order master equation
%   [Eq.~\eqref{eq:mastereq} in the main text], while dots show the results given by the EoMs 
%   [Eqs.~\eqref{eq:EoMgamma} and \eqref{eq:EoMlambda} in the main text]. The coupling   constants are all different and they have a relative standard deviation   of 32\% for the first ensemble, and 51\% for the second. The remaining parameters have values $\mu=0.5/\sqrt{N_1}$ and $\nu=0.8\mu$. For comparison we also plot the dynamics for an homogeneous system with the same $\mu$, $\nu$, and $N_{1,2}$. \label{fig:N5_inhom}}
% \end{figure}

\section{Shell model}
We define the ``shell'' $\mathcal{S}^j_\alpha$ as the 
set of nuclei $i\in\mathrm{QD}j$ with the same coupling constant 
$g_{ij} = g_{\alpha j}$. We will use Greek indices to label different 
shells. As explained in the main text, the correlations are the same for all the nuclei inside a given shell: 
$\gamma^{jj}_{ii}=x^j_\alpha$ for all $i\in\mathcal{S}^j_\alpha$; 
$\gamma^{jj'}_{ii'}=y^{jj'}_{\alpha\beta}$ for all 
$i\in\mathcal{S}^j_\alpha, i'\in\mathcal{S}^{j'}_\beta$, 
$(i,j)\neq(i',j')$; and $\lambda^{jj'}_{ii'}=z^{jj'}_{\alpha\beta}$ for all $i\in\mathcal{S}^j_\alpha, i'\in\mathcal{S}^{j'}_\beta$, 
$(i,j)\neq(i',j')$. Unless the coupling constants in a given QD are all 
different, the number of shells is smaller than the number of nuclei in 
that QD ($\alpha\leq N_j$). Thus, we can reduce the number of variables we need to consider in order to compute the polarizations and transverse spin variance. From Eqs.~(\ref{eq:gammaii}--\ref{eq:lambdaij}) (and the 
equivalent EoM for QD2), using the Z-rules approximation, we obtain the 
following set of equations:
\begin{align}
  \Delta x^j_\alpha & = -(-1)^j \kappa g_{\alpha j} 
  \big(2 S^j_\alpha + g_{\alpha j}\big)
  -\eta g_{\alpha j}^2 \big(2 x^j_\alpha - 1\big) \,, \label{eq:x}
  \\[1em]
  \begin{split}
    \Delta y^{jj}_{\alpha\beta} & = (-1)^j \kappa 
    \Big[g_{\beta j} \big(2x^j_\beta - 1\big) S^j_{\alpha\beta}
    + g_{\alpha j} \big(2x^j_\alpha - 1\big) S^j_{\beta\alpha} 
    + g_{\alpha j} g_{\beta j} \big(x^j_\alpha + x^j_\beta - 1\big)\Big] \\
    & \quad + 2 \kappa \big(g_{\alpha j} S^j_{\alpha\beta} 
    + g_{\beta j} S^j_{\beta\alpha}\big) y^{jj}_{\alpha\beta}
    - \eta \big(g_{\alpha j}^2 + g_{\beta j}^2\big) y^{jj}_{\alpha\beta} 
    + \eta g_{\alpha j} g_{\beta j} z^{j}_{\alpha\beta} \,,
  \end{split}
  \\[1em]
  \begin{split}
    \Delta z^{jj}_{\alpha\beta} & = - (-1)^j 2\kappa 
    \Big[g_{\alpha j} \big(2x^j_\beta - 1\big) 
    \big(2S^j_{\alpha\beta} + g_{\alpha j}\big) 
    + g_{\beta j} \big(2x^j_\alpha - 1\big) 
    \big(2S^j_{\beta\alpha} + g_{\beta j}\big)\Big] \\
    & \quad - 8\kappa \big(g_{\beta j} S^j_{\alpha\beta} 
    + g_{\alpha j} S^j_{\beta\alpha}\big) y^{jj}_{\alpha\beta}
    + 8\eta g_{\alpha j} g_{\beta j} y^{jj}_{\alpha\beta} 
    - 2\eta \big(g_{\alpha j}^2 + g_{\beta j}^2\big) z^{jj}_{\alpha\beta} 
    \,. \label{eq:z}
  \end{split}
\end{align}
Here, we have defined $S^j_\alpha = \sum_\gamma g_{\gamma j} N^j_\gamma 
y^{jj}_{\alpha\gamma} - g_{\alpha j} y^{jj}_{\alpha\alpha}$, and
$S^j_{\alpha\beta} = S^j_\alpha - g_{\beta j} y^{jj}_{\alpha\beta}$; 
$N^j_\alpha$ denotes the number of nuclei in shell $\mathcal{S}^j_\alpha$.
Note that it only makes sense to consider $y^{jj}_{\alpha\alpha}$ and 
$z^{jj}_{\alpha\alpha}$ if $N^j_\alpha > 1$, otherwise we should set
$y^{jj}_{\alpha\alpha} = z^{jj}_{\alpha\alpha} = 0$. As for variables
involving shells in different QDs, we obtain form Eqs.~(\ref{eq:gamma12}, 
\ref{eq:lambda12}), using the Z-rules approximation, the following 
equations:
\begin{align}
  \begin{split}
    \Delta y^{12}_{\alpha\beta} & = - \kappa g_{\alpha 1} 
    \left[(2 x^1_\alpha - 1) A_{\alpha\beta}
    - 2 y^{12}_{\alpha\beta} S^1_\alpha\right]
    + \kappa g_{\beta 2} \left[(2 x^2_\beta - 1) B_{\alpha\beta}
    + 2 y^{12}_{\alpha\beta} S^2_\beta\right]\\
    & \quad - \eta (g_{\alpha 1}^2 + g_{\beta 2}^2) y^{12}_{\alpha\beta} 
    + \frac{\mu\nu}{4} g_{\alpha 1} g_{\beta 2} z^{12}_{\alpha\beta}\,,
  \end{split} \label{eq:y12} \\[1em]
  \begin{split}
    \Delta z^{12}_{\alpha\beta} & = 4\kappa g_{\alpha 1} 
    \left[(2 x^2_\beta - 1) S^1_\alpha
    + 2 y^{12}_{\alpha\beta} A_{\alpha\beta}\right]
    - 4\kappa g_{\beta 2} \left[(2 x^1_\alpha - 1) S^2_\beta
    - 2 y^{12}_{\alpha\beta} B_{\alpha\beta}\right]\\
    & \quad - 2\eta (g_{\alpha 1}^2 + g_{\beta 2}^2) z^{12}_{\alpha\beta} 
    + 2 \mu\nu g_{\alpha 1} g_{\beta 2} y^{12}_{\alpha\beta}
    - 2\kappa \left[g_{\beta 2}^2 (2 x^1_\alpha - 1) 
    - g_{\alpha 1}^2 (2 x^2_\beta - 1)\right],
  \end{split} \label{eq:z12}
\end{align}
where we have defined $A_{\alpha\beta} = \sum_\gamma g_{\gamma 1}N^1_\gamma
y^{12}_{\gamma \beta} - g_{\alpha 1} y^{12}_{\alpha\beta}$, and 
$B_{\alpha\beta} = \sum_\gamma g_{\gamma 2}N^2_\gamma
y^{12}_{\alpha \gamma} - g_{\beta 2} y^{12}_{\alpha\beta}$.

The steady state is a solution of $\Delta x^j_\alpha = 0$, 
$\Delta y^{jj'}_{\alpha\beta} = 0$, and $\Delta z^{jj'}_{\alpha\beta} = 0$.
Looking at this set of equations, we realize that 
$\Delta z^{jj}_{\alpha\alpha} = -4\Delta y^{jj}_{\alpha\alpha}$, so the 
system is underdetermined. Nonetheless, at any given step $n$,
\begin{equation}
  z^{jj}_{\alpha\alpha}(n+1) = z^{jj}_{\alpha\alpha}(n) 
  + \Delta z^{jj}_{\alpha\alpha}(n) = z^{jj}_{\alpha\alpha}(n) 
  - 4\Delta y^{jj}_{\alpha\alpha}(n) \,,
\end{equation}
so the steady state solution satisfies
$z^{jj}_{\alpha\alpha} = z^{jj}_{\alpha\alpha}(0) - 4\left[
  y^{jj}_{\alpha\alpha} - y^{jj}_{\alpha\alpha}(0)\right]$, where 
$y^{jj}_{\alpha\alpha}(0)$ and $z^{jj}_{\alpha\alpha}(0)$ are the initial 
values of the corresponding variables. In this manner, the steady state 
actually depends on the initial state. Furthermore, we find numerically 
that for our particular choice of initial state, the steady state is such 
that $y^{jj}_{\alpha\beta} = 0$, for $\alpha \neq \beta$. 
Substituting in back in Eqs.~(\ref{eq:x}--\ref{eq:z}), we obtain
\begin{align}
  & 2 x^j_\alpha - 1 = -(-1)^j \frac{\kappa}{\eta} \left[2(N^j_\alpha - 1)
  y^{jj}_{\alpha\alpha} + 1\right]\,, \\
  & 4\kappa (N^j_\alpha - 2) \left[\kappa (N^j_\alpha - 1) - \eta\right] 
  \left(y^{jj}_{\alpha\alpha}\right)^2
  + \left[2\kappa^2 (2 N^j_\alpha - 3) + 6 \eta^2\right] 
  y^{jj}_{\alpha\alpha} + \kappa^2 - \eta^2 = 0\,.
\end{align}
Thus, the steady-state polarization of each shell is independent of the 
others, and it can be found solving a single quadratic equation. It depends on the values of $\mu$, $\nu$, and the number of nuclei in the shell $N^j_\alpha$, but not on its coupling constant $g_{\alpha j}$.

Having found the steady state values of $x^j_\alpha$ and $y^{jj}_{\alpha\beta}$, we can substitute them back in Eqs.~(\ref{eq:y12}, \ref{eq:z12}) and use a numerical algorithm to find the steady state values of $y^{12}_{\alpha\beta}$ (we used a version of the Levenberg-Marquardt 
algorithm).

\begin{figure}[!htb]
  \includegraphics{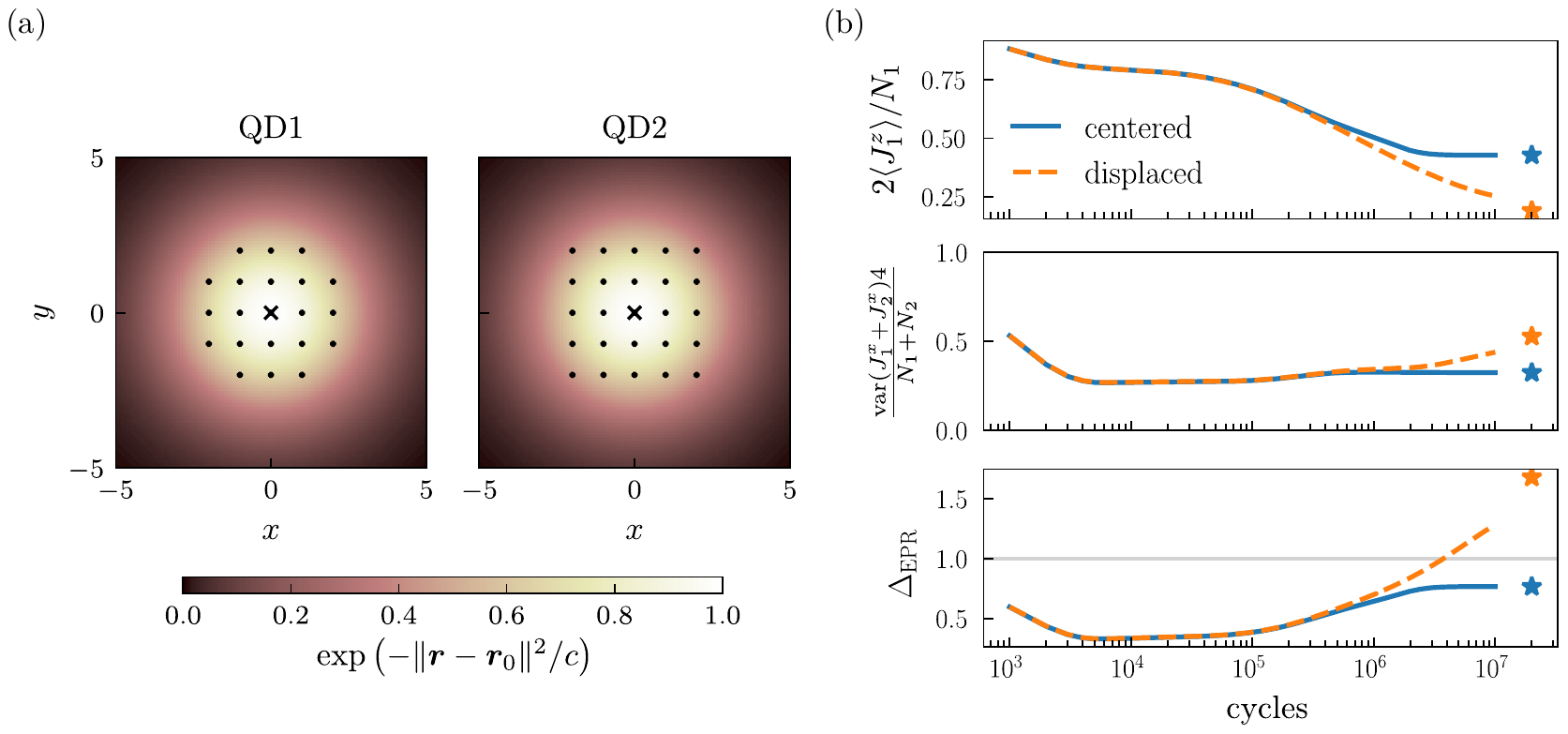}
  \caption{(a) Schematics of the system under consideration. Each ensemble
  has a total of $N_1=21$ and $N_2=25$ nuclei, arranged in a lattice as
  shown by the black dots, with coupling constants proportional to a 
  Gaussian function (colormap). The center of the Gaussian is indicated by a black cross. The case shown corresponds to the ``centered'' sytem.
  (b) Comparison between the dynamics of such system in case the Gaussian function is centered with respect to the lattice of nuclei, $\bm{r}_{0,j} = (0, 0)$ ($j=1,2$) and in case it is shifted a 
  little bit away from the center in QD1, $\bm{r}_{0,1} = (0.2, 0.1)$, 
  $\bm{r}_{0,2} = (0, 0)$. In both cases the spread of the function is the
  same $c=10$. The stars on the right side of the plot mark the steady 
  state values of the corresponding quantities. The rest of parameters are
  in both cases $\mu=0.5/\sqrt{N_1}$ and $\nu=0.8\mu$, and the initial 
  state is $\rho_0 = \ket{\Psi_\mathrm{FP}} \bra{\Psi_\mathrm{FP}}$.
  \label{fig:evolution_shifted}}
\end{figure}

As an example, in Fig.~\ref{fig:evolution_shifted} we show the dynamics of a system where we model the electronic wavefunction with a Gaussian. The two ensembles consist of $N_1=21$ and $N_2=25$ nuclei arranged in a square lattice. In one of the cases considered, the electronic wavefunction in each QD is centered with respect to the nuclear lattice, so that many nuclei have the same coupling constant. Therefore, this case can be described with few shells containing several nuclei each. In the other case, the electronic wavefunction in QD1 has the same shape, but it has been displaced a little bit away from the center of the nuclear lattice, resulting in a system with many more shells, and with fewer nuclei per shell. The former case features a higher value of the nuclear polarizations, a lower value of the transverse spin variances, and a value of $\Delta_\mathrm{EPR}<1$ in the steady state, which is reached within the computed time span. The second case, on the other hand, does not present an entangled steady state but metastable entanglement for many cycles.

\begin{figure}[!htb]
  \centering
  \includegraphics{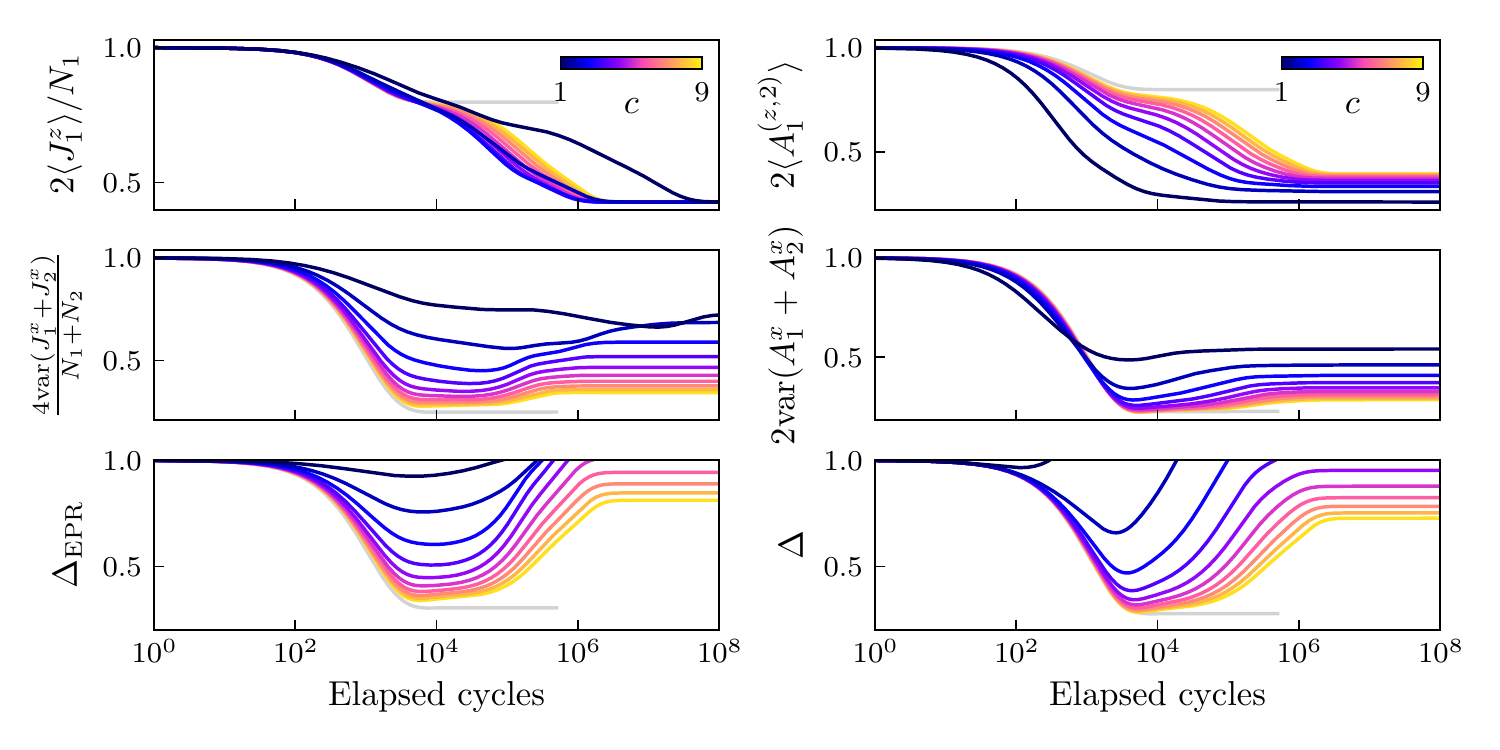}
  \caption{Comparison between the two entanglement witness $\Delta$ and 
  $\Delta_\mathrm{EPR}$ for the same system as in Fig.~\ref{fig:evolshells}
  of the main text. \label{fig:comparisonDelta}}
\end{figure}

In Fig.~\ref{fig:comparisonDelta}, we compare the two entanglement witnesses $\Delta$ and $\Delta_\mathrm{EPR}$ for the system analyzed in Fig.~\ref{fig:evolshells} in the main text. We also show the evolution of some of the quantitites needed to compute them, which present interesting differences. The polarization $\mean{J^z_1}$ reaches a fixed steady state value, the same for all the different cases considered. This is consistent with the fact that the steady state polarization of each shell is independent of the couplings. By contrast, in $\mean{A^{(z,2)}_1}$ the shell polarizations are weighted by their respective couplings squared, such that its steady state value is different in every case considered. Also, notice how the curves for the more inhomogeneous cases seem to reach the steady state value much faster than the homologous curves for $\mean{J^z_1}$. This suggests that the more weakly coupled spins are 
the ones which take longer to stabilize. As we can see, $\Delta$ detects more cases than $\EPR$ to be entangled in the steady state.

\end{document}